\newif\ifdouble
\doubletrue

\ifdouble
    \documentclass[journal, 10pt, twoside]{IEEEtran}
\else
    \documentclass[journal, 12pt, draftcls, onecolumn, twoside]{IEEEtran}
\fi

\usepackage{cite}
\ifCLASSINFOpdf
\usepackage[pdftex]{graphicx}
\graphicspath{{./diagrams/}{./images/}}
\DeclareGraphicsExtensions{.pdf,.jpeg,.png,.eps}
\usepackage{epstopdf}
\else
\fi
\usepackage{amsmath}
\usepackage{algorithmic}
\usepackage{algorithm}

\makeatletter
\newcommand\fs@norules{\def\@fs@cfont{\bfseries}\let\@fs@capt\floatc@ruled
	\def\@fs@pre{}%
	\def\@fs@post{}%
	\def\@fs@mid{\kern3pt}%
	\let\@fs@iftopcapt\iftrue}
\makeatother
\restylefloat{algorithm}

\usepackage{array}
\usepackage[caption=false,font=footnotesize]{subfig}
\usepackage{url}
\usepackage{balance}

\usepackage{color}
\usepackage{amsfonts}
\usepackage{amssymb}
\usepackage{tikz}
\usepackage{hyperref}
\usetikzlibrary{patterns}
\usetikzlibrary{backgrounds}
\usetikzlibrary{calc}
\usetikzlibrary{shapes}
\usetikzlibrary{fit}
\usetikzlibrary{chains}
\usetikzlibrary{arrows}
\usepackage{multirow}
\usepackage{tkz-graph}
\usetikzlibrary{intersections}
\usepackage{tkz-euclide}
\usetkzobj{all}
\usepackage{multirow}
\usepackage{tkz-graph}

\tikzstyle{block} = [draw, rectangle, minimum height=2em, minimum width=1em]
\tikzstyle{sum} = [draw, circle, node distance=1cm, fill=white]
\tikzstyle{input} = [coordinate] \tikzstyle{output} = [coordinate]
\tikzstyle{pinstyle} = [pin edge={to-,thin,black}]

\usepackage[normalem]{ulem}

\makeatletter
\let\cite\relax
\DeclareRobustCommand{\cite}{%
  \let\new@cite@pre\@gobble
  \@ifnextchar[\new@cite{\@citex[]}}
\def\new@cite[#1]{\@ifnextchar[{\new@citea{#1}}{\@citex[#1]}}
\def\new@citea#1{\def\new@cite@pre{#1}\@citex}
\def\@cite#1#2{[{\new@cite@pre\space#1\if\relax\detokenize{#2}\relax\else, #2\fi}]}
\makeatother

\begin{document}
\title{A Framework for Iterative Frequency Domain EP-based Receiver Design}

 \author{Serdar~\c{S}ah\.{i}n,
		Antonio~Maria~Cipriano,
		Charly~Poulliat
		and~Marie-Laure~Boucheret%
		%Michael~Shell,~\IEEEmembership{Member,~IEEE,}
        %John~Doe,~\IEEEmembership{Fellow,~OSA,}
        %and~Jane~Doe,~\IEEEmembership{Life~Fellow,~IEEE}% <-this % stops a space
\thanks{Draft for submission to IEEE Trans. on Commun., manuscript received April 6, 2018, revised June 26, 2018. Results related to this paper will also be presented, in part, at EUSIPCO'18, and at PIMRC'18.}
\thanks{S. \c{S}ah\.{i}n is with both Thales Communications \& Security and IRIT/INPT-ENSEEIHT (e-mail: serdar.sahin@thalesgroup.com}%
\thanks{A. M. Cipriano is with Thales Communications \& Security, 4 Av. des Louvresses, 92230, Gennevilliers, France (e-mail: antonio.cipriano@thalesgroup.com)}%
\thanks{C. Poulliat and M.-L. Boucheret are with IRIT/INPT-ENSEEIHT, 2 Rue Charles Camichel, 31000, Toulouse, France (e-mails: charly.pouillat@enseeiht.fr, marie-laure.boucheret@enseeiht.fr)}%
}
%\thanks{M. Shell was with the Department
%of Electrical and Computer Engineering, Georgia Institute of Technology, Atlanta,


\markboth{Draft for IEEE Journal on Transactions on Communications - June 2018}%
{\c{S}ah\.{i}n \MakeLowercase{\textit{et al.}} -  A Framework for Iterative Frequency Domain EP-based Receiver Design}

	\IEEEpubid{
		\begin{minipage}{\textwidth}\ \\[12pt] \centering
			\copyright IEEE. Personal use of this material is permitted. However, permission to reprint/republish this material for advertising or promotional purposes or for creating new collective works for resale or redistribution to servers or lists, or to reuse any copyrighted component of this work in other works must be obtained from the IEEE.
		\end{minipage}
	}


\maketitle

\begin{abstract}
An original expectation propagation (EP) based message passing framework is introduced, wherein transmitted symbols are considered to belong to the multivariate \emph{white} Gaussian distribution family. 
This approach allows deriving a novel class of single-tap frequency domain (FD) receivers with a quasi-linear computational complexity in block length, thanks to Fast-Fourier transform (FFT) based implementation.
This framework is exposed in detail, through the design of a novel double-loop single-carrier frequency domain equalizer (SC-FDE), where \emph{self-iterations} of the equalizer with the demapper, and \emph{turbo iterations} with the decoder, provide numerous combinations for the performance and complexity trade-off. 
Furthermore, the flexibility of this framework is illustrated with the derivation of an overlap FDE, used for time-varying channel equalization, among others, and with the design of a FD multiple-input multiple-output (MIMO) detector, used for spatial multiplexing. 
Through these different receiver design problems, this framework is shown to improve the mitigation of inter-symbol, inter-block and multi-antenna interferences, compared to alternative single-tap FD structures of previous works.
Thanks to finite-length and asymptotic analysis, supported by numerical results, the improvement brought by the proposed structures is assessed, and then completed by also accounting for computational costs.
\end{abstract}


% \fi
\IEEEpeerreviewmaketitle

\section{Introduction}
%\subsection{Motivations}

\ifdouble
    \IEEEPARstart{N}EXT generation wireless communication systems require sophisticated interference mitigation techniques to meet the ever-increasing demands for improved throughput despite being limited in frequency and time resources \cite{higuchi_2015_noma_SIC}.
\else
    Next generation wireless communication systems require sophisticated interference mitigation techniques to meet the ever-increasing demands for improved throughput despite being limited in frequency and time resources \cite{higuchi_2015_noma_SIC}.
\fi
Moreover, computationally-efficient frequency domain (FD) receivers are of interest for cellular or wireless ad hoc networks where low-cost radios are involved. 
For instance, Long Term Evolution (LTE) uplink, device-to-device and vehicle-to-vehicle communications in 4G 3GPP, and its evolutions, use single-carrier (SC) or single-carrier frequency division multiple access (SC-FDMA) waveforms with frequency domain equalizers (FDE) to mitigate inter-symbol interference (ISI) in quasi-static wideband channels \cite{bazzi1117_V2V_long}.
 
From a communication theory perspective, the design of receivers with affordable complexity to reach the optimum maximum likelihood joint detection and decoding performance is of interest. 
A major milestone, in this regard, is the discovery of turbo-codes, which paved the way for research on iterative processing techniques built around soft-input soft-output (SISO) receivers \cite{bauch_1994_comparisonSISOturbodetection}.
In particular, a turbo receiver using a maximum a posteriori (MAP) detector and a MAP decoder is able to operate at channel symmetric information rate (SIR), thanks to the BCJR algorithm, at the expense of a exponentially scaling computational complexity \cite{douillard_iterative_1995,bahl_optimal_1974}.

\IEEEpubidadjcol
In equalization, where MAP detectors are limited to applications with low modulation orders and very short delay spreads, a vast literature exists on extending conventional minimum mean square error (MMSE) linear equalizer (LE) or decision feedback equalizer (DFE) to turbo processing with interference cancellation (IC) \cite{glavieux_turbo_1997,wang_1999_iterativeSoftICCDMA,tuchler_minimum_2002,tuchler_turbo_2011}.
Among those, block receivers, %built with filter-banks, 
offer best performance with a computational cost scaling at best quadratically in block length, and approximate finite-impulse response receivers have quadratic complexity in channel spread.

When the statistics of the prior symbol feedback from the decoder is white (i.e. the reliability of prior estimates is static over the block), block linear equalizers (BLE) can be efficiently implemented via FFTs as FD LE, with the so-called "one-tap" filters, where each frequency bin (also called sub-carrier) is independently processed in parallel. 
Hence, in general, by whitening the estimates used for IC, iterative receivers can be built using one-tap FDEs, with a computational complexity scaling quasi-linearly with the block length \cite{tuchler_linear_2001, tuchler_turbo_2011}.

Despite the improvements brought by the turbo-iterations, there is a significant gap between FDE achievable rates and the channel SIR, especially in moderately or highly selective channels. 
Consequently, non-linear extensions have been explored to improve FDE performance \cite{falconer_frequency_2002, benvenuto_comparison_2002,koppler_2003_combinedfrequency,benvenuto_2005_iterative,ng_turbofrequency_2007,grossman_2008_nonlinearFDE_PDA,tao_2015_singlecarrierfreq}.

    \begin{table*}[t!]
    	\caption{Developments Related to Iterative Equalization on Single-tap Frequency Domain Equalization.}
    	\label{table:refclass}
    	\centering
    	\begin{tabular}{|c||l||c||c||c|}
    		\hline
    		\multirow{2}{*}{Reference} & \multirow{2}{*}{Contribution on single-tap FDE} & Decoder & Decision & \multirow{2}{*}{Schedule}\\  
    		 & & Feedback &  Feedback & \\  \hline \hline
    		1973 \cite{walzmanSchwartz_73_automaticEqualizationFDE} & Initially proposed frequency domain equalization & - & -  & - \\ \hline
    		\multirow{2}{*}{1994 \cite{sari_frequencydomain_1994}} & Revived FDE through the comparison of SC-FDE relative to & \multirow{2}{*}{-}  & \multirow{2}{*}{-}  & \multirow{2}{*}{-} \\ 
    		 & multicarrier signalling. & & & \\ \hline
    		2001 \cite{tuchler_linear_2001} & Derived the turbo iterated FD LE-IC. Denoted as FD LE-EXTIC. & Extrinsic & - & Parallel/- \\  \hline
    		2002 \cite{falconer_frequency_2002} & A hybrid DFE with a frequency domain feedforward filter and & \multirow{2}{*}{-} & \multirow{2}{*}{Hard} &  \multirow{2}{*}{-/Serial}  \\
    		2002 \cite{benvenuto_comparison_2002} & a time domain feedback filter is proposed. & & &  \\ \hline
    		2003 \cite{koppler_2003_combinedfrequency} & Simplifies hybrid DFE design by using noise prediction. & - & Hard &  -/Serial \\ \hline
    		\multirow{2}{*}{2005 \cite{benvenuto_2005_iterative}} & Proposed a non-linear receiver with FD feedforward and feedback & \multirow{2}{*}{-} & \multirow{2}{*}{Hard/APP} &  \multirow{2}{*}{-/Parallel} \\ 
    		 &filters, called iterative block DFE (IBDFE). & & & \\ \hline
    		\multirow{2}{*}{2006 \cite{visoz_frequency-domain_2006}} & Proposed a turbo FD MIMO receiver. It used APP estimates from & \multirow{2}{*}{APP} & \multirow{2}{*}{-} & \multirow{2}{*}{Parallel/-} \\
    	 	& the decoder instead of extrinsic. Denoted as FD LE-APPIC. & & & \\ \hline
    		\multirow{2}{*}{2007 \cite{ng_turbofrequency_2007}} & Compared FDE with FD feedforward and TD/FD feedback filters.  & \multirow{2}{*}{Extrinsic} & \multirow{2}{*}{Hard/APP} & Parallel/- \\
    		&  TD/FD are equivalent (parallel schedule). Soft better than hard.    & & & -/Parallel \\ \hline
    		\multirow{2}{*}{2008 \cite{grossman_2008_nonlinearFDE_PDA}} & Proposed a BPSK receiver that is self-iterated with APP estimates & \multirow{2}{*}{Extrinsic} &  \multirow{2}{*}{APP} & \multirow{2}{*}{Par./Par.}\\
    		 & before decoding at each turbo iteration. & & & \\ \hline
    		\multirow{2}{*}{{2013 \cite{guo_2013_iterativeFDE_GAMPext}}} & {Derived a self-iterated turbo receiver based on GAMP, that exploits} & \multirow{2}{*}{{Extrinsic}} &  \multirow{2}{*}{{APP}} & \multirow{2}{*}{{Par./Par.}}\\
    		 & {APP estimates at each turbo iteration.} & & & \\ \hline
    		2015 \cite{chenZheng_2015_fdte_equivalence} & Equivalence of coded IBDFE to FD LE-EXTIC is shown. & - & - & - \\ \hline
    		\multirow{2}{*}{2015 \cite{tao_2015_singlecarrierfreq}} & Extended results in \cite{tuchler_linear_2001, benvenuto_2005_iterative, grossman_2008_nonlinearFDE_PDA} to a turbo FDE with a APP &  \multirow{2}{*}{Extrinsic} & \multirow{2}{*}{APP} & \multirow{2}{*}{Par./Par.}\\
    		& -based self-iteration, denoted as FD SILE-APPIC. & & & \\ \hline
    		\textbf{This Paper} & Proposes a self-iterated FD LE-IC with EP-based feedback. 
    		& Extrinsic & Extrinsic & Par./Par. \\ \hline
    	\end{tabular}
    \end{table*}

Recently, new ideas on Bayesian inference, used in the field of artificial intelligence for solving classification or probability density functions (PDF) estimation problems arouse the interest of the communication theory and signal processing communities. Expectation propagation \cite{minka_2001_expectationpropagation} is a technique for approximate Bayesian inference, which can be used as a message passing algorithm  that extends the loopy belief propagation (BP) \cite{minka2005divergence}, conventionally used for turbo receiver design. Indeed, EP is used with variables having PDF from the exponential family, which allows for the computation of symbol-wise extrinsic information, in the context of soft demapping, that was lost when using BP \cite{senstAscheid_2011_frameworkEP_MMSEMIMO}. There are various recent receiver proposals, that observed remarkable performance improvements by exploiting EP \cite{walsh_2006_iterativedecodingwithEP, senstAscheid_2011_frameworkEP_MMSEMIMO, sun_2015_combinedBPEP_kalmanSmoother, santosMurilloFuentes_2017_EPBLE, santosMurilloFuentes_2017_EPnuBLE,sahinCipriano_2018_DFEEP}.

This paper introduces a novel category of frequency domain receivers, obtained by a specific framework of expectation propagation based message passing algorithm.
This approach is exposed through the design of an elementary FDE framework, and then the impact of this methodology on more advanced receivers is shown through equalization of time-varying channels with overlap FDE, and with spatial multiplexing with FD multi-antenna detectors. Results on the use of this approach for SC-FDE design are partially exposed in \cite{sahin_2018_IterativeEqEP_FDE}, and extension of results herein to SC-FDMA multi-user detectors is exposed in \cite{sahin_2018_SpectrallyMUDEP}.

\subsection{Related Work}

There is a significant amount of work on iterative equalization which would deserve a survey paper on its own, here we restrict ourselves to the significant developments in prior work related to single-tap FDEs, and to EP-based receivers.

\subsubsection{Iterative Single-Tap FD Receivers}

There is a long research track on frequency domain equalizers, starting from very low-complexity linear FDE up to non-linear FD turbo equalizers. Table \ref{table:refclass} lists chronological milestones on developments regarding how interference cancellation with either decoder or decision (demapper) feedback is used. The position of the FDE derived in Section \ref{sec:fdsile} of this paper is also shown. The ``schedule" column indicates in which manner the decoder/demapping feedback is used by the equalizer.
 
First FD turbo linear equalizer - interference canceller (FD LE-IC) was derived using conventional turbo formalism \cite{wang_1999_iterativeSoftICCDMA,tuchler_minimum_2002} to yield the extrinsic (EXT) feedback based FD LE-EXTIC \cite{tuchler_linear_2001}. 
However, as \cite{witzke2002iterative} noted, using a posteriori probability (APP) based feedback from the decoder yields significant improvement in turbo detection. Turbo FDE was extended to FD LE-APPIC \cite{visoz_frequency-domain_2006}. Nevertheless, APP feedback violates the independence principle of turbo iterative systems \cite{douillard_iterative_1995}, so theoretical background for such structures was absent.

Independently of the emerging turbo equalization literature, given that time domain (TD) block DFE structures outperformed block LE \cite{kaleh_1995_channelequalization}, derivation of non-linear FDE was of interest. In particular, a hybrid implementation of block DFE was carried out in \cite{falconer_frequency_2002, benvenuto_comparison_2002}. This structure uses a FD feedforward filter and a TD feedback filterbank, which carried out symbol-wise, i.e. serial, interference cancellation with hard decisions. The use of noise prediction in \cite{koppler_2003_combinedfrequency}, simplified the computation of hybrid DFE, by forcing the feedforward filter to be the same as the FD LE filter, while the overall structure remained equivalent to block DFE.

In \cite{benvenuto_2005_iterative}, the frequency domain feedback concept was introduced, and denoted iterative block DFE (IBDFE). This structure uses decision feedback in a blockwise, parallel schedule, allowing the use of FFTs over feedback symbol block, and significantly reducing complexity. Despite its name, this structure is a LE-IC, with the decision feedback being used for interference cancellation, and it is not related to the TD block DFE in \cite{kaleh_1995_channelequalization}. 
Indeed, the TD block DFE of \cite{kaleh_1995_channelequalization} uses serial symbol-wise hard decision feedback via a fairly complicated feedback filterbank, and thus it is unrelated to the linear IC scheme of \cite{benvenuto_2005_iterative}.
In \cite{ng_turbofrequency_2007}, variations of IBDFE were evaluated with hard or soft APP, and TD or FD feedback. It is noted in \cite{ng_turbofrequency_2007, benvenuto_2010_singlecarrier} that when used with forward error correction, this structure is equivalent to FD LE-EXTIC. 

In \cite{grossman_2008_nonlinearFDE_PDA}, probabilitic data association is used to derive a non-linear FDE for BPSK, through a self-iterated MMSE LE-IC using APP feedback from previous detections, before computing extrinsic LLRs for decoding. This structure, and IBDFE \cite{benvenuto_2005_iterative,ng_turbofrequency_2007} were later extended to generalized constellations in \cite{tao_2015_singlecarrierfreq}. In the latter work, non-linear block FDE (similar complexity to block LE) were evaluated, using APP decision feedback with serial and parallel schedules. These results are then used to derive a single-tap FD self-iterated LE-IC with an initial IC carried out with EXT feedback from the decoder, followed by a second round of IC carried out with APP feedback from the detector. Here, this structure is denoted as FD SILE-APPIC. 
{Another APP-based iterative FDE is derived via generalized approximate message passing (GAMP) \cite{guo_2013_iterativeFDE_GAMPext}.}

\subsubsection{Receivers based on Expectation Propagation}

As stated in the introductory paragraphs, EP reignited interest in digital receiver design thanks to a novel type of soft symbol estimates, computed at the demapper, which respects the independence principle of turbo iterative systems, unlike soft APP estimates. 
EP paradigm has been used for iterative multiple-input multiple-output (MIMO) detection \cite{senstAscheid_2011_frameworkEP_MMSEMIMO}, for time domain iterative equalization with Kalman smoothers \cite{sun_2015_combinedBPEP_kalmanSmoother}, with block LE \cite{santosMurilloFuentes_2017_EPBLE}, with filter LE \cite{santosMurilloFuentes_2017_EPnuBLE} or with filter DFE \cite{sahinCipriano_2018_DFEEP}.
Most receivers listed above exploit EP through self-iterations, allowing the demapper to compute an extrinsic feedback for IC. As demodulation is cheaper than decoding in computational costs, self-iterations provide alternative performance-complexity trade-offs to turbo-iterations.

Expectation propagation has been used in frequency domain in \cite{zhang_2016_expectationPropMIMOGFDM, wu_2017_spectralefficientbandallocation} mainly for the mitigation of inter-band interference. The former reference uses it for a generalized frequency division multiplexing receiver, as an iterative block receiver, with cubic complexity in block length, and with a single self-iteration. That structure is extended for SC-FDMA in \cite{wu_2017_spectralefficientbandallocation} under the acronym of joint-EP (J-EP). The latter reference also includes a single-tap simplification of that receiver, denoted distributed-EP (D-EP), which was however obtained through a zero-forcing type derivation, which makes it severely vulnerable to spectral nulls \cite[eq. (48)]{wu_2017_spectralefficientbandallocation}.

In this paper, instead of using EP on symbols distributed in multivariate Gaussian distributions, as in \cite{senstAscheid_2011_frameworkEP_MMSEMIMO, santosMurilloFuentes_2017_EPBLE,santosMurilloFuentes_2017_EPnuBLE,zhang_2016_expectationPropMIMOGFDM,wu_2017_spectralefficientbandallocation}, it uses EP with \emph{white} multivariate Gaussian distributions.

\subsection{Contributions and Paper Outline}

This paper's contributions are novel receivers which ensue from an EP-based message passing framework, where transmitted symbols are assumed to belong to the multivariate \emph{white} Gaussian distributions. 
These structures use single-tap FD MMSE linear filters, with interference cancellation using the EP-based extrinsic feedback, and they are shown to outperform alternatives from the previous works. Moreover the complexity of these structures have quasi-linear dependence on the block length, unlike cubic or quadratic dependencies of EP-based receivers in prior work.
{Proposed approach is also compared with approximate message passing (AMP) algorithms that are used in various estimation problems such as compressed sensing \cite{rangan_2010_GAMP, rangan_2016_VAMP}.}

This approach, which has not been previously used for digital receiver design to the authors' knowledge, is exposed with the design and analysis of a FDE for quasi-static frequency-selective channels.
Then to give a glimpse of the full potential of this framework, more advanced receivers such as an overlap FDE, or a FD MIMO detector are derived and evaluated.

The remainder of this paper is organized as follows. Section \ref{sec:fdsile} presents SC-FDE receiver design with the proposed message passing framework, and the resulting receiver is analyzed in section \ref{sec:fdeCompare}. In section \ref{sec:overlapFDE}, the application of this framework is considered for time-varying channel equalization via an overlap FDE, and in section \ref{sec:fdMIMO}, a multi-antenna spatial multiplexing application is considered. Conclusions are drawn in the end.

\subsection{Notations}

Bold lowercase letters are used for vectors: let $\mathbf{u}$ be a $N \times 1$ vector, then $u_n, n=0,\dots,N-1$ are its entries. 
Capital bold letters denote matrices: for a given $N \times M$ matrix $\mathbf{A}$, $[\mathbf{A}]_{n,:}$ and $[\mathbf{A}]_{:,m}$ respectively denote its $n^\text{th}$ row and $m^\text{th}$ column, and $a_{n,m}=[\mathbf{A}]_{n,m}$ is the entry $(n,m)$. Underlined vector $\underline{\mathbf{x}}$ denotes the frequency domain representation of $\mathbf{x}$.

$\mathbf{I}_N$ is the $N\times N$ identity matrix, $\mathbf{0}_{N,M}$ and $\mathbf{1}_{N,M}$ are respectively all zeros and all ones $N\times M$ matrices. $\mathbf{e}_n$ is the $N\times 1$ indicator whose only non-zero entry is $e_n=1$.
Operator $\textbf{Diag}(\mathbf{u})$ denotes the diagonal matrix whose diagonal is defined by $\mathbf{u}$.
$\mathbb{R}, \mathbb{C}$, and $\mathbb{F}_k$ are respectively the real field, the complex field and a Galois field of order $k$.
Let $x$ and $y$ be two random variables, then $\mu_x=\mathbb{E}[x]$ is the expected value, $\sigma_x^2=\text{Var}[x]$ is the variance and $\sigma_{x,y}=\text{Cov}[x,y]$ is the covariance. The probability of $x$ taking a value $\alpha$ is $\mathbb{P}[x=\alpha]$, and probability density functions (PDF) are denoted as $p(\cdot)$.
If $\mathbf{x}$ and $\mathbf{y}$ are random vectors, then we define vectors $\pmb{\mu}_\mathbf{x}=\mathbb{E}[\mathbf{x}]$ and $\pmb{\sigma}_\mathbf{x}^2=\text{Var}[\mathbf{x}]$, the covariance matrix $\mathbf{\Sigma}_{\mathbf{x},\mathbf{y}}=\textbf{Cov}[\mathbf{x,y}]$ and we note $\mathbf{\Sigma}_{\mathbf{x}}=\textbf{Cov}[\mathbf{x,x}]$.
$\mathcal{CN}(\mu_x,\sigma_x^2)$ denotes the circularly-symmetric complex Gaussian distribution of mean $\mu_x$ and variance $\sigma_x^2$, and  $\mathcal{B}(p)$ denotes the Bernoulli distribution with a success probability of $0\leq p \leq 1$.

\section{Proposed EP-based Design Framework for a BICM SC-FDE System}\label{sec:modelling}
\label{sec:fdsile}

\subsection{System Model}

    Single-carrier transmission of a block of $K$ symbols using a bit-interleaved coded modulation (BICM) scheme is considered.  
    In detail, the information block $\mathbf{b} \in \mathbb{F}_2^{K_b}$ is encoded and then interleaved into a codeword $\mathbf{d} \in \mathbb{F}_2^{K_d}$, with a code rate $R_c=K_b/K_d$. A memoryless modulator $\varphi$ maps $\mathbf{d}$ into $\mathbf{x} \in \mathcal{X}^K$, where the constellation $\mathcal{X}$ has $M$ elements, and where $K=K_d/q$, with $q=\log_2 M$. This constellation is assumed to have a zero mean, and average power $\sigma_x^2=1$, with equiprobable symbols. This operation associates the $q$-word $\mathbf{d}_{k}\triangleq[d_{qk},\dots,d_{q(k+1)-1}]$ to the symbol $x_{k}$, and $\varphi^{-1}_j(x_{k})$ and $d_{k,j}$ are used to refer to $d_{kq+j}$.

    An equivalent baseband circular channel model is considered, including the effects of transceiver modules and the channel propagation. The receiver is assumed to be ideally synchronized in time and frequency and it has perfect channel state information. 
    The received samples are given by
    \begin{equation}
        \mathbf{y} = \mathbf{H}\mathbf{x} + \mathbf{w}, \label{eq_chan_td}
    \end{equation}
    where $\mathbf{H} \in \mathbb{C}^{K \times K}$ is a circulant matrix, generated by $\mathbf{h} = \left[h_{0},\dots,h_{L-1}, \mathbf{0}^T_{K-L,1}\right]$, the impulse response extended with $K-L$ zeros, $L<K$ being the channel spread. Unlike the additive white Gaussian noise (AWGN) model in \cite{sahin_2018_IterativeEqEP_FDE}, here, a coloured and correlated noise $\mathbf{w}$ is considered to capture the impact of eventual interfering signals, with $\mathbf{{w}} \sim \mathcal{CN}(\mathbf{0}_K, \mathbf{{\Sigma}}_{\mathbf{w}})$. 
    This channel model is applicable to different SC-FDE implementations such as the cyclic prefix (CP) SC-FDE, or the zero-padded (ZP) SC-FDE  \cite{muquet_cyclic_2002}, among others.

    The normalized $K$-DFT matrix is given by its elements $[\mathcal{F}_K]_{m,n}$ $= \exp(-2j\pi nm/K)/\sqrt{K}$, such that $\mathcal{F}_K\mathcal{F}_K^H = \mathbf{I}_K$. 
    Then the equivalent frequency domain transmission model is
    \begin{equation}
        \mathbf{\underline{y}} = \mathbf{\underline{H}}\mathbf{\underline{x}} + \mathbf{\underline{w}}, \label{eq_circmodel}
    \end{equation}
    with $\mathbf{\underline{x}}=\mathcal{F}_K\mathbf{x}$, $\mathbf{\underline{y}}=\mathcal{F}_K\mathbf{y}$, $\mathbf{\underline{w}}=\mathcal{F}_K\mathbf{w}$, and $\mathbf{\underline{H}}=\mathcal{F}_K\mathbf{H}\mathcal{F}_K^H=\textbf{Diag}(\mathbf{\underline{h}})$ with the channel frequency response being
    \begin{equation}
        \underline{h}_{k} = {\textstyle\sum}_{l=0}^{L-1} h_{l} \exp(-2j\pi kl/K), \quad k=1,\dots,K,
    \end{equation} 
    and $\mathbf{\underline{w}} \sim \mathcal{CN}(\mathbf{0}_K, \mathbf{\underline{\Sigma}}_{\mathbf{w}})$, with $\mathbf{\underline{\Sigma}}_{\mathbf{w}} =\mathcal{F}_K \mathbf{\Sigma}_{\mathbf{w}} \mathcal{F}_K^H$ is the noise covariance matrix in the FD. To keep the receiver complexity low, the non-diagonal elements of  $\mathbf{\underline{\Sigma}}_{\mathbf{w}}$ are ignored, hence use-cases involving interference with non-negligible inter-carrier correlations are out of scope.

The remainder of this section covers the approximation of the posterior probability density function of transmitted symbols, by using an EP-based message passing algorithm. In particular, symbol variables $\mathbf{x}$ are assumed to belong to a multivariate white Gaussian distribution, of the form $\mathcal{CN}(\mathbf{\bar{x}}, \bar{v}\mathbf{I}_K)$, where the reliability of symbol estimates $\mathbf{\bar{x}}$ is given by a scalar $\bar{v}$. The resulting approximate distribution is shown to yield a novel iterative single-tap FD LE-IC.

\subsection{Factor Graph Model}

The joint posterior probability density function (PDF) of data bits, given FD observations, is $p(\mathbf{b},\mathbf{d},\mathbf{x} \vert \mathbf{\underline{y}})$.
The optimal joint MAP receiver operating on FD observations resolves the criterion $\hat{\mathbf{b}}=\max_\mathbf{b} p(\mathbf{b} \vert \mathbf{\underline{y}})$. Assuming i.i.d. information bits, the posterior PDF is factorized as
\begin{equation}
p(\mathbf{b}, \mathbf{d}, \mathbf{x} \vert \mathbf{\underline{y}}) \propto \underbrace{p(\mathbf{\underline{y}}\vert\mathbf{x})}_\text{channel} \underbrace{p(\mathbf{x}\vert \mathbf{d})}_\text{mapping} \underbrace{p(\mathbf{d\vert b})}_\text{encoding}.\label{eq_postb}
\end{equation}
This density is further factorized by using the memoryless mapping $p(\mathbf{x}\vert \mathbf{d}) = \prod_{k=0}^{K-1}p( x_k\vert \mathbf{d}_k)$, and the independence assumption in BICM encoding $p(\mathbf{d\vert b}) = \prod_{k=0}^{K-1}\prod_{j=0}^{q-1}p(d_{k,j})$, where the probability mass function (PMF) $p(d_{k,j})~\triangleq~p(d_{k,j}\vert \mathbf{b})$ is seen as a Bernoulli-distributed prior constraint provided by the decoder, from the receiver's point of view. 

\begin{figure}[!t]
	\centering
	\begin{tikzpicture}[thick, xscale=0.85, yscale=0.85, every node/.style={xscale=0.85, yscale=0.85}]
	% Latent node
	\tikzstyle{latent} = [circle,fill=white,draw=black,inner sep=0.15pt,
	minimum size=12pt, node distance=0]
	% Observed node
	\tikzstyle{obs} = [latent,fill=gray!25]
	% Constant node
	\tikzstyle{const} = [rectangle, inner sep=0pt, node distance=1]
	% Factor node
	\tikzstyle{factor} = [rectangle, fill=black,minimum size=8pt, inner
	sep=0pt, node distance=0]
	
	\node[obs, label=90:$\mathbf{\underline{y}}$] at (-0.25, 3) (yk){};

	\node[factor] at (0.75, 3) (qequ){};
	
	\node[latent, label=180:$x_{0}$] at (1.75, 2) (x1){};
	\node at (1.75, 2.6) {$\vdots$};
	\node[latent, label=45:$x_k$] at (1.75, 3) (xk){};
	\node at (1.75, 3.6) {$\vdots$};
	\node[latent, label=180:$x_{K-1}$] at (1.75, 4) (xend){};
	
	\node at (2.5, 2) (qdem1){};
	\node[factor] at (3.75, 3) (qdemk){};
	\node at (2.5, 4) (qdemend){};

	\node[latent, label=270:$d_{k,0}$] at (5.25, 2.5) (ckq){};
	\node at (5.25, 3.1) (cqkdots) {$\vdots$};
	\node[latent, label=90:$d_{k,q-1}$] at (5.25, 3.5) (cqkp1){};

	\node[factor] at (6.25, 2.5) (qdec_kq){};
	\node at (6.25, 3.1) (qdecdots) {$\vdots$};
	\node[factor] at (6.25, 3.5) (qdec_kqp1){};
	
	\draw (0.5, 2.5) [dashed, rounded corners=2mm]rectangle(1, 3.5);
	\node at (0.1, 2.2) (qequt){$f_\text{EQU}(\mathbf{x})$};
	\draw (3.5, 2.5) [dashed, rounded corners=2mm]rectangle(4, 3.5);
	\node[label=below:$f_\text{DEM}(x_k\text{, }\mathbf{d}_k)$] at (3.75, 2.5) (qdemt){};
	\draw (6, 2) [dashed, rounded corners=2mm]rectangle(6.5, 4);
	\node[label=below:$f_\text{DEC}(\mathbf{d}_k)$] at (7.25, 3.5) (qdect){};

	\draw [-] (yk) -- (qequ);

	\draw [-] (qequ) -- (x1);
	\draw [-] (qequ) -- (xk);
	\draw [-] (qequ) -- (xend);

	\draw [-] (ckq) -- (qdemk);
	\draw [-] (cqkp1) -- (qdemk);
	
	\draw [-, dotted] (qdem1) -- (x1);
	\draw [-] (qdemk) -- (xk);
	\draw [-, dotted] (qdemend) -- (xend);

	\draw (ckq) -- (qdec_kq);
	\draw (cqkp1) -- (qdec_kqp1);

	\end{tikzpicture}
	\caption{Factor graph for the posterior (\ref{eq_pdfFactors}) on $x_k$ and $\mathbf{d}_k$.}
	\label{fig_fg_rec}
\end{figure}
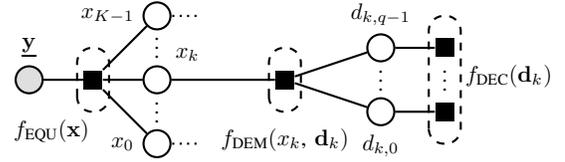

As we focus on iterative detection and decoding for a given transmission, we will focus on the posterior for the estimation of variables $d_{k,j}$ and $x_k$, and remove $\mathbf{b}$ from notations. Hence (\ref{eq_postb}) can be factorized as
    \begin{align}
		p(\mathbf{d},\mathbf{x}  \vert \mathbf{\underline{y}})
		&\propto \textstyle p(\mathbf{\underline{y}} \vert \mathbf{x})\prod_{k=0}^{K-1}p(x_{k}\vert \mathbf{d}_{k})\prod_{j=0}^{q-1} p(d_{k,j}). \label{eq_pdfFactors}
	\end{align}
This process is iteratively carried out by a message-passing based detection and decoding algorithm operating on the variables nodes (VN) $x_k$ and $d_{k,j}$ by using constraints imposed by factor nodes (FN) corresponding to the factorization of the posterior PDF in (\ref{eq_pdfFactors}). 
The equalization (EQU) FN resolves the multipath channel constraints with 
	\begin{equation}
    	\label{eq_equ_fn}
		f_\text{EQU}(\mathbf{x}) \triangleq  p(\mathbf{\underline{y}} \vert \mathbf{x}) \propto e^{-\mathbf{\underline{y}}^H\mathbf{\underline{\Sigma}}_{\mathbf{w}}^{-1}\mathbf{\underline{y}}+2\mathcal{R}(\mathbf{\underline{y}}^H\mathbf{\underline{\Sigma}}_{\mathbf{w}}^{-1}\mathbf{\underline{H}}\mathcal{F}_{K}\mathbf{x})},
	\end{equation}
where the dependence on $\mathbf{\underline{y}}$ is omitted, as the FD observations are unchanged during iterative detection. 
Demapper (DEM) FN handles the mapping constraints with
	\begin{equation}
		f_\text{DEM}(x_{k}, \mathbf{d}_{k}) \triangleq p(x_{k}\vert \mathbf{d}_{k}) \propto \textstyle\prod_{j=0}^{q-1} \delta(d_{k,j}-\varphi^{-1}_{j}(x_{k})), \label{eq_dem_fn}
	\end{equation}
where $\delta$ is the Dirac delta function, and finally, channel coding constraints are handled by
	\begin{equation}
		{f}_\text{DEC}(\mathbf{d}_{k}) \triangleq  \textstyle\prod_{j=0}^{q-1} p(d_{k,j}).
	\end{equation}
The considered BICM SC-FDE system factor graph is given by Fig. \ref{fig_fg_rec}. Note that, unlike the finite-impulse response receiver factor graph in \cite{sahinCipriano_2018_DFEEP}, EQU FN impacts all transmitted symbols.

\subsection{Proposed EP-based Message Passing Framework with White Gaussian Distributions}

Expectation propagation, extends belief propagation as a message passing algorithm by assuming the variable nodes to have PDFs belonging to the exponential family \cite{minka2005divergence}. This results in exchanged messages to be depicted by tractable distributions, which allows for the iterative computing of a fully-factorized approximation of challenging PDFs such as $p(\mathbf{d}, \mathbf{x}  \vert \mathbf{\underline{y}})$. Resulting approximation can then be marginalized on variables of interest, to yield the desired estimates.

Updates involving a FN F, connected to variable nodes $\mathbf{v}$ are as follows.
Messages exchanged between VN $v_i$, the $i^\text{th}$ component of $\mathbf{v}$, and $\text{F}$ are given by
    \begin{eqnarray}
        m_{v\rightarrow \text{F}}(v_i) &\triangleq&  \textstyle\prod_{\text{G} \neq F} m_{\text{G} \rightarrow v}(v_i), \label{eq_MESSfromVN} \\
        m_{\text{F}\rightarrow v}(v_i) &\triangleq& {\text{proj}_{\mathcal{Q}_{v_i}} \left[q_\text{F}(v_i)\right]}/{m_{v\rightarrow \text{F}}(v_i)},\label{eq_MESSfromFN}
    \end{eqnarray}
where $\text{proj}_{\mathcal{Q}_{v_i}}$ is the Kullback-Leibler projection towards the probability distribution $\mathcal{Q}_{v_i}$ of VN $v_i$. The approximate posterior $q_\text{F}(v_i)$ is an estimation of the marginal of the true posterior $p(\mathbf{v})$ on $v_i$, obtained by combining the true factor on FN F with messages from the neighbouring VNs
    \begin{equation}
		q_\text{F}(v_i) \triangleq \textstyle\int_{\mathbf{v}^{\backslash i}} f_\text{F}(\mathbf{v})\prod_{v_j} m_{v \rightarrow \text{F}}(v_j) \mathbf{dv}^{\backslash i},  \label{eq_postFN}
	\end{equation}
where $\mathbf{v}^{\backslash i}$ are VNs without $v_i$ \cite{minka2005divergence}.
The projection operation for exponential families is equivalent to \emph{moment matching}, which simplifies the computation of messages \cite{minka2005divergence}.

For the proposed framework, our simplifying assumption is that VNs $\mathbf{x}$ lie in multivariate white Gaussian distributions. 
Hence, a message involving these VNs is fully characterized by a \emph{vector mean} and a \emph{scalar variance}. 
On the other hand $d_{k,j}$ follow a Bernoulli distribution (which is included in the exponential family), whose messages are characterized by binary log-likelihood ratios (LLRs), as in the conventional belief propagation algorithm.

\subsection{Derivation of Exchanged Messages}

In this subsection, the framework above is applied to the considered factor graph, by first, defining exchanged messages, and then computing their characterizing parameters. 

The messages arriving on the VN $x_k$ are Gaussians with
\begin{eqnarray}
    m_{\text{EQU}\rightarrow x}(x_{k}) \propto \mathcal{CN}\left( x_k^e, v^e \right), \label{eq_equ_to_x} \\
    m_{\text{DEM}\rightarrow x}(x_{k}) \propto \mathcal{CN}\left( x_k^d, v^d\right), \label{eq_dem_to_x}
\end{eqnarray}
where means are dependent on $k$ and variances are static. 
Oppositely, the messages arriving on the VN $d_{k,j}$ are Bernoullis
\begin{eqnarray}
    m_{\text{DEC}\rightarrow d}(d_{k,j}) \propto \mathcal{B}\left( p_d^a \right),\,
    m_{\text{DEM}\rightarrow d}(d_{k,j}) \propto \mathcal{B}\left( p_d^e \right). \label{eq_FN_to_d}
\end{eqnarray}
The features, i.e the characteristic parameters, of these distributions are updated following a selected schedule, during the message passing procedure. 
For Bernoulli distributions, it is rather preferable to work with bit LLRs, rather than the success probability $p_d$:
\begin{equation}
L(d_j) \triangleq \ln\frac{\mathbb{P}[d_j=0]}{\mathbb{P}[d_j=1]} = \ln\frac{1-p_d}{p_d}.
\end{equation}
We use $L_a(\cdot)$, $L_e(\cdot)$ and $L(\cdot)$ operators to denote respectively a priori, extrinsic and a posteriori LLRs. When applied to $d_{k,j}$, this vocabulary represents the SISO receiver's perspective, i.e. $L_a(d_{k,j})$, $L_e(d_{k,j})$ respectively characterize $m_{\text{DEC}\rightarrow d}(d_{k,j})$ and $m_{\text{DEM}\rightarrow d}(d_{k,j})$. 
Fig.\ref{fig_sim} illustrates a conventional view of the receiver with the quantities above.

Finally, considering the factor graph shown on Fig. \ref{fig_fg_rec}, all variable nodes are only connected to a pair of distinct factor nodes. Consequently, using eq. (\ref{eq_MESSfromVN}), 
$m_{v \rightarrow \text{F}}(v_i) = m_{\text{G}\rightarrow v}(v_i)$, 
for all VN $v_i$, and FN $\text{F}, \text{G},\, \text{F}\neq\text{G}$ they are connected to.

\ifdouble
    % double: TURBO EQUALIZATION DIAGRAM
    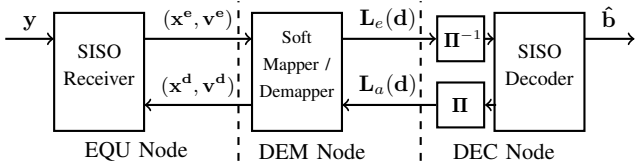
\begin{figure}[!t]
    	\centering
    	\begin{tikzpicture}[thick, scale=0.85, every node/.style={scale=0.85}]
    
        \draw [-, dashed] (1.7,-1.5) -- (1.7,1);
        \draw [-, dashed] (-1.15,-1.5) -- (-1.15,1);
        \node at (-2.75,-1.35) () {EQU Node};
        \node at (0,-1.35) () {DEM Node};
        \node at (3,-1.35) () {DEC Node};
    
    	% NODES DEFINITION
    	\matrix [ row sep = 0.05cm, column sep = 0.15cm, cells={scale=1.0}]
    	{
    		%\\
    		%& &
    		%\node (ch_est)[yshift = 0cm, xshift=-0.42cm] {$\mathbf{H}$}; 
    		%\node (snr_est)[yshift = 0cm, xshift=0.42cm] {$ \sigma_w^2$}; &
    		%& & & & & %& 
    		%& &&
    		%\\
    		% —————————— row 1
    		\node (ch_input)[xshift=-0cm] {}; &
    		\node (eq_input)[yshift = 0.25cm, xshift=-0cm] {$\mathbf{y}$};
    		&
    		\node (dummy_equ_in1) [coordinate, xshift=-0.15cm] {};
    		\node (dummy_equ1) [xshift=-0cm, yshift=0.3cm] {};&&&&&
    		\node (dummy_equ_out1) [coordinate, xshift=0.15cm] {};
    		&
    		\node (eqoutput)[above]{\small $(\mathbf{{x}^e},\mathbf{v^e})$}; 
    		&
    		\node (dummy_dem_in1) [coordinate, xshift=-0.15cm] {};
    		\node (dummy_dem1) [xshift=-0cm, yshift=0.3cm] {};&&&&&
    		\node (dummy_dem_out1) [coordinate, xshift=0.15cm] {};
    		%\node (demap) [block, fill=white] {\parbox[c]{1.4cm}{\centering \footnotesize Soft \\Demapper}}; 
    		&
    		\node (llrx)[above]{$\mathbf{L}_e(\mathbf{d})$}; &
    		\node (deintrlv) [block, fill=white] {\parbox[c]{.5cm}{\small \centering $\pmb{\Pi}^{-1}$}}; 
    		&
    		%\node (llrc)[above]{$\mathbf{L}_a(\mathbf{c})$}; &
    		\node (dummy_dec_in1) [coordinate, xshift=-0.15cm] {};
    		\node (dummy_dec1) [xshift=-0cm, yshift=0.3cm] {};&&&&&
    		\node (dummy_dec_out1) [coordinate, xshift=0.15cm] {};
    		%\node (decode) [block] {\parbox[c]{1.4cm}{\centering \small SISO \\ Decoder}}; 
    		&
    		\node (dec_out) [yshift= 0.35cm]{$\mathbf{\hat{b}}$};
    		&
    		\node (dec_output)[coordinate] {};
    		\\
    		%\\
    		%&&
            %&
    		%\node (x_mean_d)[yshift = 0.25cm] {\small $(\mathbf{{x}^d},\mathbf{v^d})$};
    		%\node (x_dec)[coordinate] {};&
    		%&&&&&&
    		%\\
    		%\\
    		% —————————— row 2
    		&&
    		&&&&&
    		\node (dummy_equ_in2) [coordinate, xshift=0.15cm] {};
    		\node (dummy_equ2) [xshift=0cm, yshift=-0.3cm] {};
    		&
    		\node (x_mean)[above] {\small $(\mathbf{{x}^d},\mathbf{v^d})$};
    		&
    		\node (dummy_dem_out2) [coordinate, xshift=-0.15cm] {};
    		&&&&&
    		\node (dummy_dem_in2) [coordinate, xshift=0.15cm] {};
    		\node (dummy_dem2) [xshift=0cm, yshift=-0.3cm] {};
    		%\node (map) [block, fill=white] {\parbox[c]{1.4cm}{\centering \small Soft \\ Mapper}}; 
    		&
    		\node (llrxe)[above]{$\mathbf{L}_a(\mathbf{d})$}; &
    		\node (intrlv) [block, fill=white] {\parbox[c]{.5cm}{\small \centering $\pmb{\Pi}$}}; 
    		&
    		%\node (llrce)[above]{$\mathbf{L}_e(\mathbf{c})$}; 
    		\node (dummy_dec_out2) [coordinate, xshift=-0.12cm] {};
    		&&&&&
    		\node (dummy_dec_in2) [coordinate, xshift=0.12cm] {};
    		\node (dummy_dec2) [xshift=0cm, yshift=-0.3cm] {};
    		&&
    		\\
    	};
    	\node (equalizer) [fit=(dummy_equ1)(dummy_equ2), block] {\centering \small  SISO Receiver};
    	\node (demap) [fit=(dummy_dem1)(dummy_dem2), block] {\centering \footnotesize Soft Mapper / Demapper};
    	\node (decode) [fit=(dummy_dec1)(dummy_dec2), block] {\centering \small SISO Decoder};
    
    	% PATHS
    	\draw [->] (ch_input) -- (dummy_equ_in1);
    	%\draw [->] (ch_est) -- (equalizer.127);
    	%\draw [->] (snr_est) -- (equalizer.53);
    	\draw [->] (dummy_equ_out1) -- (dummy_dem_in1);
    	%\draw [->] (demap.270) |- (x_dec) -| (equalizer.300);
    	\draw [->] (dummy_dem_out1) -- (deintrlv);
    	\draw [->] (deintrlv) -- (dummy_dec_in1);
    	\draw [->] (dummy_dec_out1) -- (dec_output);
    	\draw [->] (dummy_dec_out2) -- (intrlv);
    	\draw [->] (intrlv) -- (dummy_dem_in2);
    	%\draw [->] (map.180) -| (equalizer.300);
    	\draw [->] (dummy_dem_out2) -- (dummy_equ_in2);
    	
    	% auxiliary nodes
    	% \node [coordinate, xshift = 0.4cm, yshift = 1cm] (nAux1) at (n26) {};
    	% \node [coordinate, xshift = -0.4cm, yshift = -1cm] (nAux2) at (n12) {};
    	% \draw [dashed] (nAux1) -| (nAux2) -| (nAux1) node [above, pos = 0.38] {feedback linearized system};
    	\end{tikzpicture}
    	\caption{Factor nodes shown as an iterative BICM receiver.}
    	\label{fig_sim}
    \end{figure}
\else
    % single: TURBO EQUALIZATION DIAGRAM
    \begin{figure}[!t]
    	\centering
    	\begin{tikzpicture}[thick, scale=0.85, every node/.style={scale=0.85}]
    
        \draw [-, dashed] (2.2,-2) -- (2.2,1.5);
        \draw [-, dashed] (-1.45,-2) -- (-1.45,1.5);
        \node at (-3.25,-1.75) () {EQU Node};
        \node at (0,-1.75) () {DEM Node};
        \node at (3.75,-1.75) () {DEC Node};
    
    	% NODES DEFINITION
    	\matrix [ row sep = 0.05cm, column sep = 0.2cm, cells={scale=1.0}]
    	{
    		%\\
    		%& &
    		%\node (ch_est)[yshift = 0cm, xshift=-0.42cm] {$\mathbf{H}$}; 
    		%\node (snr_est)[yshift = 0cm, xshift=0.42cm] {$ \sigma_w^2$}; &
    		%& & & & & %& 
    		%& &&
    		%\\
    		% —————————— row 1
    		\node (ch_input)[xshift=-0cm] {}; &
    		\node (eq_input)[yshift = 0.25cm, xshift=-0cm] {$\mathbf{y}$};
    		&
    		\node (dummy_equ_in1) [coordinate, xshift=-0.15cm] {};
    		\node (dummy_equ1) [xshift=-0cm, yshift=0.3cm] {};&&&&&
    		\node (dummy_equ_out1) [coordinate, xshift=0.15cm] {};
    		&
    		\node (eqoutput)[above]{\small $(\mathbf{{x}^e},\mathbf{v^e})$}; 
    		&
    		\node (dummy_dem_in1) [coordinate, xshift=-0.15cm] {};
    		\node (dummy_dem1) [xshift=-0cm, yshift=0.3cm] {};&&&&&
    		\node (dummy_dem_out1) [coordinate, xshift=0.15cm] {};
    		%\node (demap) [block, fill=white] {\parbox[c]{1.4cm}{\centering \footnotesize Soft \\Demapper}}; 
    		&
    		\node (llrx)[above]{$\mathbf{L}_e(\mathbf{d})$}; &
    		\node (deintrlv) [block, fill=white] {\parbox[c]{.5cm}{\small \centering $\pmb{\Pi}^{-1}$}}; 
    		&
    		%\node (llrc)[above]{$\mathbf{L}_a(\mathbf{c})$}; &
    		\node (dummy_dec_in1) [coordinate, xshift=-0.15cm] {};
    		\node (dummy_dec1) [xshift=-0cm, yshift=0.3cm] {};&&&&&
    		\node (dummy_dec_out1) [coordinate, xshift=0.15cm] {};
    		%\node (decode) [block] {\parbox[c]{1.4cm}{\centering \small SISO \\ Decoder}}; 
    		&
    		\node (dec_out) [yshift= 0.35cm]{$\mathbf{\hat{b}}$};
    		&
    		\node (dec_output)[coordinate] {};
    		\\
    		%\\
    		%&&
            %&
    		%\node (x_mean_d)[yshift = 0.25cm] {\small $(\mathbf{{x}^d},\mathbf{v^d})$};
    		%\node (x_dec)[coordinate] {};&
    		%&&&&&&
    		%\\
    		%\\
    		% —————————— row 2
    		&&
    		&&&&&
    		\node (dummy_equ_in2) [coordinate, xshift=0.15cm] {};
    		\node (dummy_equ2) [xshift=0cm, yshift=-0.3cm] {};
    		&
    		\node (x_mean)[above] {\small $(\mathbf{{x}^d},\mathbf{v^d})$};
    		&
    		\node (dummy_dem_out2) [coordinate, xshift=-0.15cm] {};
    		&&&&&
    		\node (dummy_dem_in2) [coordinate, xshift=0.15cm] {};
    		\node (dummy_dem2) [xshift=0cm, yshift=-0.3cm] {};
    		%\node (map) [block, fill=white] {\parbox[c]{1.4cm}{\centering \small Soft \\ Mapper}}; 
    		&
    		\node (llrxe)[above]{$\mathbf{L}_a(\mathbf{d})$}; &
    		\node (intrlv) [block, fill=white] {\parbox[c]{.5cm}{\small \centering $\pmb{\Pi}$}}; 
    		&
    		%\node (llrce)[above]{$\mathbf{L}_e(\mathbf{c})$}; 
    		\node (dummy_dec_out2) [coordinate, xshift=-0.12cm] {};
    		&&&&&
    		\node (dummy_dec_in2) [coordinate, xshift=0.12cm] {};
    		\node (dummy_dec2) [xshift=0cm, yshift=-0.3cm] {};
    		&&
    		\\
    	};
    	\node (equalizer) [fit=(dummy_equ1)(dummy_equ2), block] {\centering \small  SISO Receiver};
    	\node (demap) [fit=(dummy_dem1)(dummy_dem2), block] {\centering \footnotesize Soft Mapper / Demapper};
    	\node (decode) [fit=(dummy_dec1)(dummy_dec2), block] {\centering \small SISO Decoder};
    
    	% PATHS
    	\draw [->] (ch_input) -- (dummy_equ_in1);
    	%\draw [->] (ch_est) -- (equalizer.127);
    	%\draw [->] (snr_est) -- (equalizer.53);
    	\draw [->] (dummy_equ_out1) -- (dummy_dem_in1);
    	%\draw [->] (demap.270) |- (x_dec) -| (equalizer.300);
    	\draw [->] (dummy_dem_out1) -- (deintrlv);
    	\draw [->] (deintrlv) -- (dummy_dec_in1);
    	\draw [->] (dummy_dec_out1) -- (dec_output);
    	\draw [->] (dummy_dec_out2) -- (intrlv);
    	\draw [->] (intrlv) -- (dummy_dem_in2);
    	%\draw [->] (map.180) -| (equalizer.300);
    	\draw [->] (dummy_dem_out2) -- (dummy_equ_in2);
    	
    	% auxiliary nodes
    	% \node [coordinate, xshift = 0.4cm, yshift = 1cm] (nAux1) at (n26) {};
    	% \node [coordinate, xshift = -0.4cm, yshift = -1cm] (nAux2) at (n12) {};
    	% \draw [dashed] (nAux1) -| (nAux2) -| (nAux1) node [above, pos = 0.38] {feedback linearized system};
    	\end{tikzpicture}
    	\caption{Factor nodes shown as an iterative BICM receiver.}
    	\label{fig_sim}
    \end{figure}
\fi

\subsubsection{Messages from DEC to DEM}

DEC FN is assumed to be a SISO channel decoder, that generates prior information $L_a(\mathbf{d})$ to DEM, when extrinsic LLRs $L_e(\mathbf{d})$ is given to it by DEM. 

Using these prior LLRs with the mapping constraints in (\ref{eq_dem_fn}), the prior PMF on $x_k=\alpha$, is
\begin{equation}
    \mathcal{P}_{k}(\alpha) \propto \textstyle\prod_{j=0}^{q-1}e^{-\varphi^{-1}_j(\alpha)L_a(d_{k,j})}, \forall \alpha\in\mathcal{X}. \label{eq_priors}
\end{equation}
This is a categorical PMF corresponding to the marginal of $f_\text{DEM}(x_{k}, \mathbf{d}_{k})m_{d\rightarrow \text{DEC}}(\mathbf{d}_{k})$ on $x_{k}$ \cite{senstAscheid_2011_frameworkEP_MMSEMIMO}, used hereafter to compute approximate marginals $q_\text{DEM}(x_k)$ and $q_\text{DEM}(d_{k,j})$.

\subsubsection{Messages from DEM to EQU}
An approximate posterior on the variable node $x_k$ is computed at the demapper, using eq. (\ref{eq_postFN}), with
\ifdouble
    % Double column
    \begin{equation}
        \begin{split}
    	    q_\text{DEM}(x_k) = \textstyle  \sum_{\mathbf{d}_k} & \textstyle  f_\text{DEM}(x_k, \mathbf{d}_k) m_{x \rightarrow \text{DEM}}(x_{k})\\
    	    & \textstyle \prod_{j=0}^{q-1} m_{d \rightarrow \text{DEM}}(d_{k,j}).
        \end{split}
    \end{equation}
\else
    % Single column
    \begin{equation}
	    q_\text{DEM}(x_k) = \textstyle   \sum_{\mathbf{d}_k} f_\text{DEM}(x_k, \mathbf{d}_k) m_{x \rightarrow \text{DEM}}(x_{k}) \prod_{j=0}^{q-1} m_{d \rightarrow \text{DEM}}(d_{k,j}).
    \end{equation}
\fi
This is a posterior categorical PMF on $x_k=\alpha$, given by eqs. (\ref{eq_equ_to_x})
and  (\ref{eq_priors}), denoted as
\begin{equation}
    \mathcal{D}_{k}(\alpha) \propto \exp{\left(-{|\alpha-x^e_{k}|^2}/{v_k^e}\right)}\mathcal{P}_{k}(\alpha),\, \forall \alpha\in\mathcal{X}. \label{eq_posteriors}
\end{equation}
For computing messages towards EQU via eq. (\ref{eq_MESSfromFN}), the posterior PMF is projected into a Gaussian distribution through moment matching. The mean and the variance of $\mathcal{D}_{k}$ are
\begin{equation}
    \label{eq_dem_stats}
    \begin{aligned}
        \mu^d_k&\triangleq\mathbb{E}_{\mathcal{D}_k}[x_k]&={}&\textstyle\sum_{\alpha\in\mathcal{X}}\alpha \mathcal{D}_k(\alpha),\\
        \gamma^d_{k}&\triangleq\text{Var}_{\mathcal{D}_k}[x_k]&={}&\textstyle\sum_{\alpha\in\mathcal{X}}\vert\alpha\vert^2 \mathcal{D}_k(\alpha) - \vert \mu_k^d \vert^2.
    \end{aligned}
\end{equation}
The result of the projection on $x_k$ is $\mathcal{CN}(\mu^d_k, \gamma^d)$, using moment matching \cite[eq. (19)]{minka2005divergence}, where means are matched, but the variance needs to satisfy, $\forall k$,
    $\gamma^d=\gamma^d_k$.
Working with white Gaussians creates an overdetermined constraint on the reliability of estimates without any exact solutions. An approximate solution, given by the ordinary least-squares, coincides with the sample average
\begin{eqnarray}
\gamma^d &\triangleq& K^{-1}{\textstyle\sum}_{k=0}^{K-1}\gamma_{k}^d. \label{eq_post_var_avg}
\end{eqnarray}

Then $m_{\text{DEM}\rightarrow x}(x_{k})$ is computed as in  (\ref{eq_MESSfromFN}), by using a Gaussian division \cite{minka_2001_expectationpropagation}, which yields
\begin{equation}
{x}^\star_k = \frac{\mu^d_k v^e - x^e_k\gamma^d} {v^e-\gamma^d},\, \text{and},\, {v}^\star = \frac{ v^e \gamma^d} {v^e-\gamma^d}. \label{eq_demap_ep_new}
\end{equation}
The major novelty in using EP lies in this expression; the computation of an extrinsic feedback to the equalizer from the demapper.
Attempting this with categorical distributions, as in BP, would completely remove $m_{x \rightarrow \text{DEM}}(x_{k})$, and the extrinsic ``feedback" to EQU would simply become prior PMF $\mathcal{P}_k$ \cite{senstAscheid_2011_frameworkEP_MMSEMIMO}, which results in a receiver equivalent to FD LE-EXTIC \cite{tuchler_linear_2001}.

The feedback produced by EP is erroneous, if the denominator in eq. (\ref{eq_demap_ep_new}) is negative, which may be caused by conflicts among the equalizer's output and the mapping constraints. In this case \cite{santosMurilloFuentes_2017_EPBLE} replaces the concerned $({x}_k^\star, v_k^\star)$ with their values from a previous iteration, and \cite{senstAscheid_2011_frameworkEP_MMSEMIMO} uses posteriors $(\mu^d_k, \gamma^d_k)$ instead. From experimentation not exposed here, the latter case is found to be more advantageous. However, unlike these references, the use of static variances greatly reduces the occurrence of $v^e \le \gamma^d$, and if it occurs, we use $\mu_k$ and $\gamma^d$ instead.

EP message passing minimizes local divergences (on marginal posteriors) in order to minimize a global divergence (full posterior).
Thus, it does not guarantee convergence and it might lock on undesirable fixed points. 
As in \cite[eq. (17)]{minka2005divergence},  a feature-based damping heuristic is used
% Double column
\begin{equation}
    \label{eq_demap_ep_damp_stats_new}
    \begin{aligned}
        v^{d(\text{next})} &=  \left[(1-\beta)/v^{\star} + \beta /\bar{v}^{d(\text{prev})}\right]^{-1}, \\
        x_{k}^{d(\text{next})} &=  {v}^{d(\text{next})}\left[(1-\beta)\frac{{x}_{k}^{\star}}{v^{\star}} + \beta \frac{{x}_{k}^{d(\text{prev})}}{{v}^{d(\text{prev})}}\right],
    \end{aligned}
\end{equation}
with tuning parameter $0 \leq \beta \leq 1$. 
{\cite{wang_2017_dopedEP} uses a linear smoother between DEM's and DEC's extrinsic estimates, which is inefficient in a self-iterated EQU-DEM schedule. Hence it is here extended to linearly smooth messages across self-iterations}
\begin{equation}
    \label{eq_demap_ep_filt_stats_new}
    \begin{aligned}
        v^{d(\text{next})} &=  (1-\beta)v^{\star} + \beta \bar{v}^{d(\text{prev})}, \\
        x_{k}^{d(\text{next})} &=  (1-\beta){{x}_{k}^{\star}} + \beta {{x}_{k}^{d(\text{prev})}}.
    \end{aligned}
\end{equation}
Although both approaches are observed to asymptotically lead to similar limits, with numerical experimentation, the feature-based approach often converges faster for the same $\beta$. However, inversions used in this approach cause numerical issues in some configurations, which makes the linear damping more preferable.

\subsubsection{Messages from EQU to DEM}
Approximate posterior on the VN $x_k$ (eq. (\ref{eq_postFN})) is given by
\ifdouble
    % Double column:
    \begin{equation}
        \begin{split}
            q_\text{EQU}(x_k) = \textstyle   \int_{\mathbf{x}^{\backslash k}} & \textstyle f_\text{EQU}(\mathbf{x}) \\
            & \textstyle \prod_{k'=0}^{K-1}m_{x \rightarrow \text{EQU}}(x_{k'}) \mathbf{dx}^{\backslash k}.
        \end{split}
    \end{equation}
\else
    % Single column:
    \begin{equation}
    	q_\text{EQU}(x_k) = \textstyle   \int_{\mathbf{x}^{\backslash k}} f_\text{EQU}(\mathbf{x}) \prod_{k'=0}^{K-1}m_{x \rightarrow \text{EQU}}(x_{k'}) \mathbf{dx}^{\backslash k}.
    \end{equation}
\fi
Denoting the integrand above as $\mathcal{CN}(\pmb{\mu}^\mathbf{e}, \mathbf{\Gamma}^\mathbf{e})$, and using eq. (\ref{eq_equ_fn}), we have
\begin{equation}
    \begin{aligned}
        \mathbf{\Gamma}^\mathbf{e} &= (\mathbf{I}_K/v^{d} + \mathcal{F}_{K}^H\mathbf{\underline{H}}^H\mathbf{\underline{\Sigma}}_{\mathbf{w}}^{-1}\mathbf{\underline{H}}\mathcal{F}_{K})^{-1}, \\
        \pmb{\mu}^\mathbf{e} &= \mathbf{\Gamma}^{\mathbf{e}} (\mathbf{x}^{d}/v^{d} + \mathcal{F}_{K}^H\mathbf{\underline{H}}^H\mathbf{\underline{\Sigma}}_{\mathbf{w}}^{-1}\mathbf{\underline{y}}),
    \end{aligned}
\end{equation}
where $\mathbf{x}^{d}=[x^{d}_0,\dots, x^{d}_{K-1}]$.
Using some matrix algebra, and Woodbury's identity on $\mathbf{\Gamma}^\mathbf{e}$, the variance $\gamma^e_k$ and the mean $\mu^e_k$ of the marginalized PDF $q_\text{EQU}(x_k)$ are given by
\begin{equation}
    \begin{aligned}
        \gamma_k^e &= \mathbf{e}_k^H{\mathbf{\Gamma}}^\mathbf{e}\mathbf{e}_k = v^{d}(1-v^{d}{\xi}), \\
        \underline{\mu}_k^e &= \mathbf{e}_k^H\mathcal{F}_{K}{\pmb{\mu}}^\mathbf{e} = \underline{x}^{d}_{k} + v^{d}{\xi}\underline{f}_{k}^{*}(\underline{y}_k-\underline{h}_k\underline{x}_k^{d}),    
    \end{aligned}
\end{equation}
where $\underline{\mu}_k^e$ and $\underline{x}_k^{d}$ are the FD spectrum of respectively $\mu_k^e$ and $x_k^d$. Parameters $\underline{f}_{k}$ and $\xi$ follow
\begin{eqnarray}
    \underline{f}_{k} &=& {\xi}^{-1}{\underline{h}_{k}}/({\underline{\sigma}_{w_{k}}^{2}+v^{d} \vert \underline{h}_{k} \vert^2}), \label{eq_filt1} \\
    {\xi} &=& {K^{-1}}\textstyle\sum_{k=0}^{K-1}\vert\underline{h}_{k}\vert^2/({\underline{\sigma}_{w_{k}}^{2}+v^{d} \vert \underline{h}_{k} \vert^2}).   \label{eq_filt_norm}
\end{eqnarray}
Noting that $\gamma_{k}^e$ does not depend on $k$, $q_\text{EQU}(x_{k})$ already belongs to the family of white multivariate Gaussians, the projection operation in (\ref{eq_MESSfromFN}) has no effect, $\gamma^{e}=\gamma_{k}^e$, $\forall k$.
Hence the Gaussian division of  $q^\text{EQU}(x_{k})$ by $m_{x\rightarrow \text{EQU}}(x_{k})$ is readily carried out to compute parameters of $m_{\text{EQU}\rightarrow x}(x_{k})$ 
\begin{align}
    \underline{x}^e_{k} &= \underline{\bar{x}}_{k}^{d} + \underline{f}_{k}^*(\underline{y}_{k} - \underline{h}_{k}\underline{\bar{x}}_{k}^{d}), \label{eq_fde}\\
    v^e &= {\xi}^{-1}-v^{d}. \label{eq_fde_outvar}
\end{align}
Note that these expressions result in the conventional MMSE FD LE-IC structure, with interference cancellation being carried out using extrinsic EP feedback $x_k^d$.

\subsubsection{Messages from DEM to DEC}
The demapper computes an approximate posterior on the VN $d_{k,j}$ using eq. (\ref{eq_postFN}) with
\ifdouble
    % Double column:
    \begin{equation}
        \begin{split}
        	q_\text{DEM}(\mathbf{d}_{k}) = \textstyle \sum_{x_k \in \mathcal{X}} 
        	& \textstyle f_\text{DEM}(x_k, \mathbf{d}_k)  m_{x \rightarrow \text{DEM}}(x_{k}) \\
        	& \textstyle \prod_{j=0}^{q-1} m_{d \rightarrow \text{DEM}}(d_{k,j}).
        \end{split}
    \end{equation}
\else
    % Single column:
    \begin{equation}
    	q_\text{DEM}(\mathbf{d}_{k}) = \textstyle \sum_{x_k \in \mathcal{X}}  f_\text{DEM}(x_k, \mathbf{d}_k) m_{x \rightarrow \text{DEM}}(x_{k}) \prod_{j=0}^{q-1} m_{d \rightarrow \text{DEM}}(d_{k,j}).
    \end{equation}
\fi
The marginalization of this posterior on $d_{k,0},\dots,d_{k,q-1}$ \cite{senstAscheid_2011_frameworkEP_MMSEMIMO}, and the division in  (\ref{eq_MESSfromFN}) is directly carried out with bit LLRs
\begin{equation}
L_e(d_{k,j}) = \ln \frac{\sum_{\alpha\in\mathcal{X}_j^0}{\mathcal{D}_k(\alpha)}}{\sum_{\alpha\in\mathcal{X}_j^1}{\mathcal{D}_k(\alpha)}} -L_a(d_{k,j}), \label{eq_demap_llr}
\end{equation}
with $\mathcal{X}_j^p=\lbrace \alpha\in\mathcal{X}: \varphi^{-1}_j(x)=p\rbrace$ where $p\in\mathbb{F}_2$.

\ifdouble
    \begin{figure}[t]
    	\centering
    	\begin{tikzpicture}[yscale=0.9,xscale=0.85, every node/.style={yscale=0.9,xscale=0.85},thick]
    	% NODES DEFINITION
    	\matrix [ row sep = 0.25cm, column sep = 0.15cm]
    	{
    		% —————————— row 1
    		\node (obs_input1)[yshift=0.2cm,xshift=0.2cm] {$\mathbf{y}$};
    		\node (input_dummy)[coordinate] {};&
    		\node (fft1) [block, fill = white] {\parbox[c]{.6cm}{\centering $\mathcal{F}_K$}}; &
    		\node (obs_input)[yshift=0.2cm,xshift=-0.5cm] {$\underline{y}_{k}$}; &
    		\node (sum1) [sum,xshift=0.1cm] {}; &
    		\node (filter1) [block, fill = white] {\parbox[c]{.6cm}{\centering $\underline{f}^*_{k}$}}; &
    		\node (sum2) [sum] {}; &
    		\node (output)[yshift=0.21cm] {${\underline{x}}^e_{k}$}; &
    		\node (ifft1) [block, fill = white]  {\parbox[c]{.6cm}{\centering $\mathcal{F}_K^H$}}; &
    		\node (outputt)[yshift=0.21cm] {$\mathbf{{x}^e}$}; &
    		\node (demap) [block, fill = white]  {\parbox[c]{1.2cm}{\centering Demap.}}; &
    		\node (output_dummy)[coordinate] {};
    		\node (output2)[yshift=0.25cm] {$L_e(\mathbf{d})$};&
    		\\
    		% —————————— row 2
    		&&
    		&
    		&
    		\node (filter3) [block, fill = white] {\parbox[c]{.6cm}{\centering $\underline{h}_{k}$}}; &
    		&
    		\node (prior_input)[yshift=0.25cm] {${{\underline{x}}}^d_{k}$};&
    		\node (fft2) [block, fill = white]  {\parbox[c]{.6cm}{\centering $\mathcal{F}_K$}}; &
    		\node (prior_input1)[yshift=0.25cm] {$\mathbf{{x}^d}$}; &
    		\node (epmod) [block, fill = white]  {\parbox[c]{1.2cm}{\centering EP Module}}; &
    		\node (prior_dummy)[coordinate] {};
    		\node (prior_input2)[yshift=0.25cm] {$L_a(\mathbf{d})$};&
    		\\
    	};
    	% AUX
    	\begin{scope}[on background layer]	
    	\node [coordinate, xshift=-1.25cm, yshift=+0.5cm](nAux1) at (filter1) {};
    	\node [coordinate, xshift=+1.1cm, yshift=-0.7cm](nAux2) at (filter3) {};
    	\node [coordinate, xshift=+0.9cm, yshift=0.5cm](nAux3) at (demap) {};
    	\node [coordinate, xshift=-0.9cm, yshift=-0.7cm](nAux4) at (epmod) {};
    	\node [coordinate, xshift=+1.1cm, yshift=-0.5cm](nAuxX) at (filter3) {};
    	\node [coordinate, xshift=-0.9cm, yshift=-0.5cm](nAuxY) at (epmod) {};
    	\node [coordinate, xshift=1.1cm, yshift=-0.5cm](nAuxU) at (filter1) {};
    	\node [coordinate, xshift=-0.9cm, yshift=-0.5cm](nAuxV) at (demap) {};
    	\draw [dashed, fill=gray, fill opacity=0.1] (nAux1) -| (nAux2) -|  (nAux1) {};
    	\draw [dashed, fill=gray, fill opacity=0.1] (nAux3) -| (nAux4) -| (nAux3) {};
    	\draw [->] (nAuxY) -- node[pos=0.85, yshift=0.2cm]{${v^d}$} (nAuxX);
    	\draw [->] (nAuxU) -- node[pos=0.15, yshift=0.2cm]{$v^e$} (nAuxV);
    	\end{scope}
    	% PATHS
    	\draw  [->] (input_dummy) -- node[pos=0.5, yshift=-0.15cm]{\scriptsize S/P}(fft1);
    	\draw  [->] (fft1) -- node[pos=0.85, yshift=0.2cm]{\small{$+$}} (sum1);
    	\draw  [->] (sum1) -- (filter1);
    	%\draw [->] (filter1) -- (filter2);
    	\draw  [->] (filter1) -- node[pos=0.6, yshift=0.2cm]{\small{$+$}} (sum2);
    	\draw [->] (sum2) -- (ifft1);
    	\draw [->] (ifft1) -- node[pos=0.4, yshift=-0.15cm]{\scriptsize P/S}(demap);
    	\draw [->] (demap) -- (output_dummy);
    	\draw  [->] (fft2) -| node[pos=0.9, xshift=-0.2cm]{\small{$+$}} (sum2);
    	\draw [->] (filter3) -| node[pos=0.9, xshift=-0.2cm]{\small{$-$}} (sum1);
    	\draw [->] (fft2) -- (filter3);
    	\draw [->] (epmod) -- node[pos=0.6, yshift=-0.15cm]{\scriptsize S/P}(fft2);
    	\draw [->] (prior_dummy) -- (epmod);
    	
    	\draw [->,transform canvas={xshift=-0.35cm}] (demap.270) -- (epmod.90);
    	\draw [->,transform canvas={xshift=0.35cm}] (epmod.90) -- (demap.270);
    	%\draw [->] (epmod.49) -- (demap.325);
    	
    	\draw [->] (-5,-0.3) -- node[pos=0.2,yshift=0.2cm]{$\mathbf{\underline{H}}$}(-2.85,-0.3);
    	\draw [->] (-5,-1) -- node[pos=0.2,yshift=0.2cm]{$\mathbf{\Sigma}_{\mathbf{w}}$}(-2.85,-1);
    	
    	\end{tikzpicture}
    	\caption{Proposed turbo FD SILE-EPIC structure.}
    	\label{fig_fdleic}
    \end{figure}
\else
    \begin{figure}[t]
    	\centering
    	\begin{tikzpicture}[yscale=0.9,xscale=0.85, every node/.style={yscale=0.9,xscale=0.85},thick]
    	% NODES DEFINITION
    	\matrix [ row sep = 0.25cm, column sep = 0.15cm]
    	{
    		% —————————— row 1
    		\node (obs_input1)[yshift=0.2cm,xshift=0.2cm] {$\mathbf{y}$};
    		\node (input_dummy)[coordinate] {};&
    		\node (fft1) [block, fill = white] {\parbox[c]{.6cm}{\centering $\mathcal{F}_K$}}; &
    		\node (obs_input)[yshift=0.2cm,xshift=-0.5cm] {$\underline{y}_{k}$}; &
    		\node (sum1) [sum,xshift=0.1cm] {}; &
    		\node (filter1) [block, fill = white] {\parbox[c]{.6cm}{\centering $\underline{f}^*_{k}$}}; &
    		\node (sum2) [sum] {}; &
    		\node (output)[yshift=0.21cm] {${\underline{x}}^e_{k}$}; &
    		\node (ifft1) [block, fill = white]  {\parbox[c]{.6cm}{\centering $\mathcal{F}_K^H$}}; &
    		\node (outputt)[yshift=0.21cm] {$\mathbf{{x}^e}$}; &
    		\node (demap) [block, fill = white]  {\parbox[c]{1.5cm}{\centering Demap.}}; &
    		\node (output_dummy)[coordinate] {};
    		\node (output2)[yshift=0.25cm] {$L_e(\mathbf{d})$};&
    		\\
    		% —————————— row 2
    		&&
    		&
    		&
    		\node (filter3) [block, fill = white] {\parbox[c]{.6cm}{\centering $\underline{h}_{k}$}}; &
    		&
    		\node (prior_input)[yshift=0.25cm] {${{\underline{x}}}^d_{k}$};&
    		\node (fft2) [block, fill = white]  {\parbox[c]{.6cm}{\centering $\mathcal{F}_K$}}; &
    		\node (prior_input1)[yshift=0.25cm] {$\mathbf{{x}^d}$}; &
    		\node (epmod) [block, fill = white]  {\parbox[c]{1.5cm}{\centering EP Module}}; &
    		\node (prior_dummy)[coordinate] {};
    		\node (prior_input2)[yshift=0.25cm] {$L_a(\mathbf{d})$};&
    		\\
    	};
    	% AUX
    	\begin{scope}[on background layer]	
        	\node [coordinate, xshift=-1.35cm, yshift=+0.5cm](nAux1) at (filter1) {};
        	\node [coordinate, xshift=+1.2cm, yshift=-0.7cm](nAux2) at (filter3) {};
        	\node [coordinate, xshift=+1.1cm, yshift=0.5cm](nAux3) at (demap) {};
        	\node [coordinate, xshift=-1.1cm, yshift=-0.7cm](nAux4) at (epmod) {};
        	\node [coordinate, xshift=+1.2cm, yshift=-0.5cm](nAuxX) at (filter3) {};
        	\node [coordinate, xshift=-1.1cm, yshift=-0.5cm](nAuxY) at (epmod) {};
        	\node [coordinate, xshift=1.2cm, yshift=-0.5cm](nAuxU) at (filter1) {};
        	\node [coordinate, xshift=-1.1cm, yshift=-0.5cm](nAuxV) at (demap) {};
        	\draw [dashed, fill=gray, fill opacity=0.1] (nAux1) -| (nAux2) -|  (nAux1) {};
        	\draw [dashed, fill=gray, fill opacity=0.1] (nAux3) -| (nAux4) -| (nAux3) {};
        	\draw [->] (nAuxY) -- node[pos=0.85, yshift=0.2cm]{${v^d}$} (nAuxX);
        	\draw [->] (nAuxU) -- node[pos=0.15, yshift=0.2cm]{$v^e$} (nAuxV);
    	\end{scope}
    	% PATHS
    	\draw  [->] (input_dummy) -- node[pos=0.5, yshift=-0.15cm]{\scriptsize S/P}(fft1);
    	\draw  [->] (fft1) -- node[pos=0.85, yshift=0.2cm]{\small{$+$}} (sum1);
    	\draw  [->] (sum1) -- (filter1);
    	%\draw [->] (filter1) -- (filter2);
    	\draw  [->] (filter1) -- node[pos=0.6, yshift=0.2cm]{\small{$+$}} (sum2);
    	\draw [->] (sum2) -- (ifft1);
    	\draw [->] (ifft1) -- node[pos=0.4, yshift=-0.15cm]{\scriptsize P/S}(demap);
    	\draw [->] (demap) -- (output_dummy);
    	\draw  [->] (fft2) -| node[pos=0.9, xshift=-0.2cm]{\small{$+$}} (sum2);
    	\draw [->] (filter3) -| node[pos=0.9, xshift=-0.2cm]{\small{$-$}} (sum1);
    	\draw [->] (fft2) -- (filter3);
    	\draw [->] (epmod) -- node[pos=0.6, yshift=-0.15cm]{\scriptsize S/P}(fft2);
    	\draw [->] (prior_dummy) -- (epmod);
    	
    	\draw [->,transform canvas={xshift=-0.35cm}] (demap.270) -- (epmod.90);
    	\draw [->,transform canvas={xshift=0.35cm}] (epmod.90) -- (demap.270);
    	%\draw [->] (epmod.49) -- (demap.325);
    	
    	\draw [->] (-5,-0.3) -- node[pos=0.2,yshift=0.2cm]{$\mathbf{\underline{H}}$}(-3.2,-0.3);
    	\draw [->] (-5,-1) -- node[pos=0.2,yshift=0.2cm]{$\mathbf{\Sigma}_{\mathbf{w}}$}(-3.2,-1);
    	
    	\end{tikzpicture}
    	\caption{Proposed turbo FD SILE-EPIC structure.}
    	\label{fig_fdleic}
    \end{figure}
\fi

\subsection{Proposed FD Self-Iterated LE-EPIC Receiver}

As the considered factor graph has cycles, it is not possible to derive a receiver algorithm with only the messages exchanged over it; a schedule for coordinating the update of variable and factor nodes is needed.

To keep the equalization complexity reasonable, a parallel scheduling across variables nodes $x_k$ is considered, in line with conventional FD LE or block LE receivers. Note that the use of a serial schedule would yield a DFE-like structure \cite{sahinCipriano_2018_DFEEP}.

To fully exploit the benefits of the feedback computed by the demapper, a flexible double-loop FDE structure is proposed.
The first loop refers to the exchange of extrinsic information between the decoder and the demapper in a turbo-iteration (TI), while the second loop refers to the message exchange in a self-iteration (SI) between the demapper and the equalizer.

Each TI $\tau = 0,\dots, \mathcal{T}$ consists of $\mathcal{S}_\tau$ SIs (may depend on $\tau$), where EQU and DEM factor nodes are updated in parallel schedule, for $s=0,\dots,\mathcal{S}_\tau$, and then the DEC factor nodes are updated with a selected SISO decoder.
To clarify this, Algorithm~\ref{algo:fdsileicep} below explicitly describes the proposed scheduling, where involved quantities are indexed by $(\tau, s)$ in the superscript.

\begin{algorithm}[t]
	\caption{Proposed FD SILE-EPIC Receiver}
	\label{algo:fdsileicep}
	\begin{algorithmic}[1]
		\renewcommand{\algorithmicrequire}{\textbf{Input}}
		\renewcommand{\algorithmicensure}{\textbf{Output}}
		\REQUIRE $\mathbf{\underline{y}}$, $\mathbf{\underline{H}}$, ${\sigma}^2_{w}$
		\STATE Initialize decoder with $L_a^{(0)}\left(\mathbf{d}_{k}\right)=0, \forall k$.
		\FOR {$\tau = 0$ to $\mathcal{T}$}
		\STATE Initialize equ. with ${\hat{x}}^{(\tau,0)}_{k}=0, \forall k$ and $\sigma_\nu^2{}^{(\tau,0)}=+\infty$.
		\STATE Use decoder's $L_a^{(\tau)}(\mathbf{d}_{k})$ to compute  $\mathcal{P}_{k}^{(\tau)}$ via (\ref{eq_priors}), $\forall k$.
		\FOR {$s = 0$ to $\mathcal{S}_\tau$}
        \STATE Use (\ref{eq_posteriors}-\ref{eq_post_var_avg}) to update demapper posteriors, $\forall k$.
		\STATE Generate soft feedback using (\ref{eq_demap_ep_new})-(\ref{eq_demap_ep_filt_stats_new}), $\forall k$.
		\STATE Compute $\bar{\xi}^{(\tau,s)}$ using (\ref{eq_filt_norm}), and, 		$\sigma_{\nu}^{2(\tau,s)}$ using (\ref{eq_fde_outvar}).
		\STATE Equalize using (\ref{eq_filt1}) and (\ref{eq_fde}), for $k=0,\dots,K-1$.
		\ENDFOR
		\STATE Provide extrinsic outputs $L^{(\tau)}_e(\mathbf{d}_{k})$ to the decoder using (\ref{eq_demap_llr}), in order to obtain next priors $L^{(\tau+1)}_a(\mathbf{d}_k), \forall k$.
		\ENDFOR
	\end{algorithmic} 
\end{algorithm}

The iterative FDE derived in this section, by applying the EP framework in the FD, with the family of white Gaussian distributions, yields the low-complexity single-tap receiver structure shown in Fig.~\ref{fig_fdleic}. 
In the next section, the behaviour of this receiver will be assessed with achievable rate analysis and comparisons with structures from the prior work.

\section{EP-based SC-FDE Performance Evaluation}\label{sec:fdeCompare}

Here, the receiver derived in the previous section is used with a fixed number of SI $\mathcal{S}_\tau = \mathcal{S}$ per TI, and it is referred as the $\mathcal{S}$-self-iterated FD LE-IC with EP (\emph{FD $\mathcal{S}$-SILE-EPIC}).

\subsection{Asymptotic Analysis}

In order to evaluate asymptotic behaviour ($\tau \rightarrow \infty$) of the proposed receiver, extrinsic information transfer (EXIT) analysis is used \cite{brink_2000_designing}. 
This consists in characterizing the behaviour of iterative SISO components with single-parameter transfer functions, by tracking the extrinsic mutual information (MI) exchanges. The behaviour of a SISO receiver is represented by the transfer function $I_E = \mathcal{T}_R(I_A, \mathbf{h}, \mathbf{\Sigma}_\mathbf{w})$ which depends on the channel parameters, with $I_A$ and $I_E$ being the MI between coded bits and respectively the a priori and extrinsic LLRs of the module.

One primordial use of EXIT analysis is performance prediction through evaluation of MI evolution. However, this entails strong assumption on the distribution of the prior LLRs of the SISO module, which cannot be met for most receivers other than MAP detectors. This issue can cause these transfer functions to be too optimistic in some cases, providing only an upper-bound on asymptotic performance.  In this regard, the accuracy of transfer functions is assessed by comparing it with actual MI trajectories, obtained with finite-length simulations.

\ifdouble
    \begin{figure}[t!]
    	\centering
    	\includegraphics[width=3.2in]{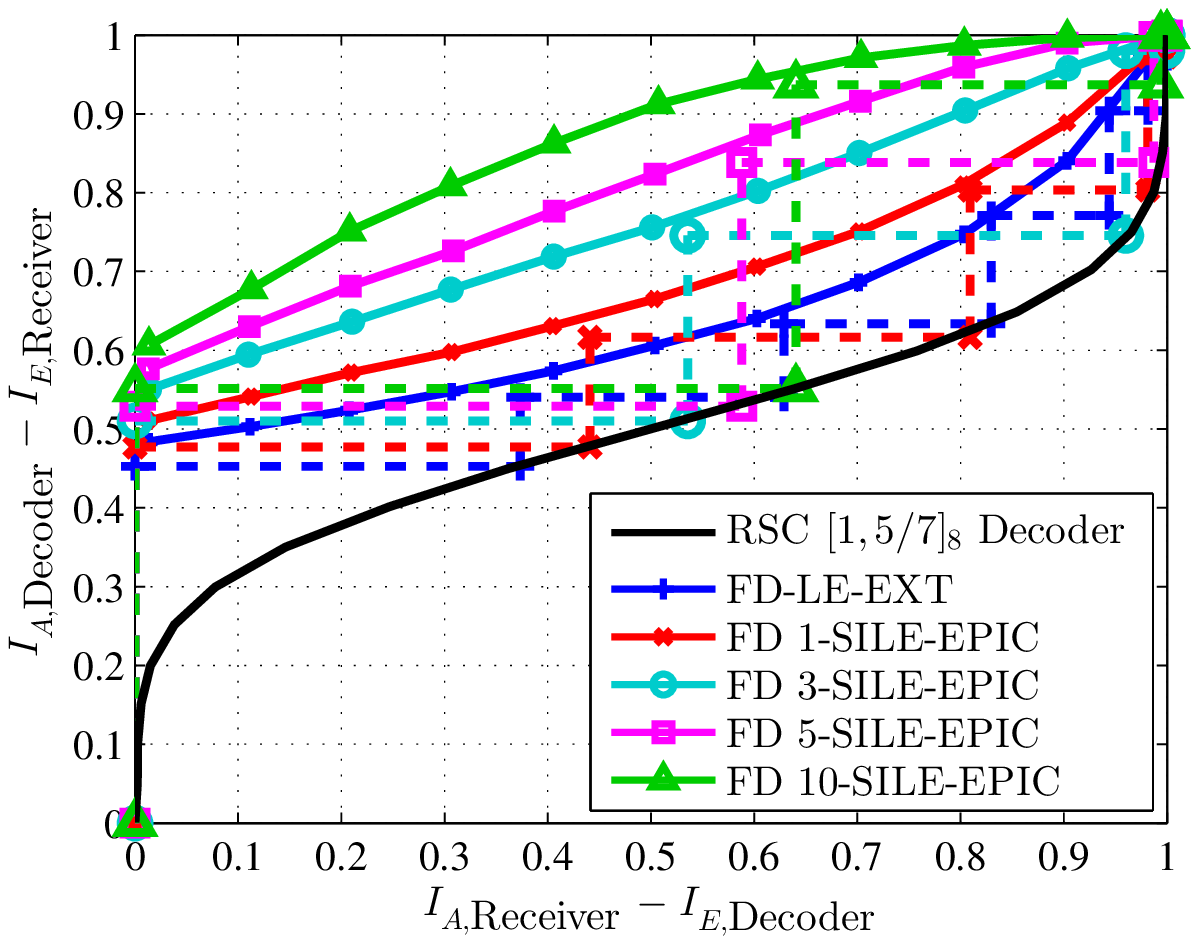}
        \caption{EXIT curves and finite-length average MI trajectories of the proposed equalizer with 8-PSK in Proakis C channel at $E_b/N_0=15$dB.}
        \label{fig:qam16ProakisC_EXITtraj}
    \end{figure}
\else
    \begin{figure}[t!]
    	\centering
    	\includegraphics[width=3.15in]{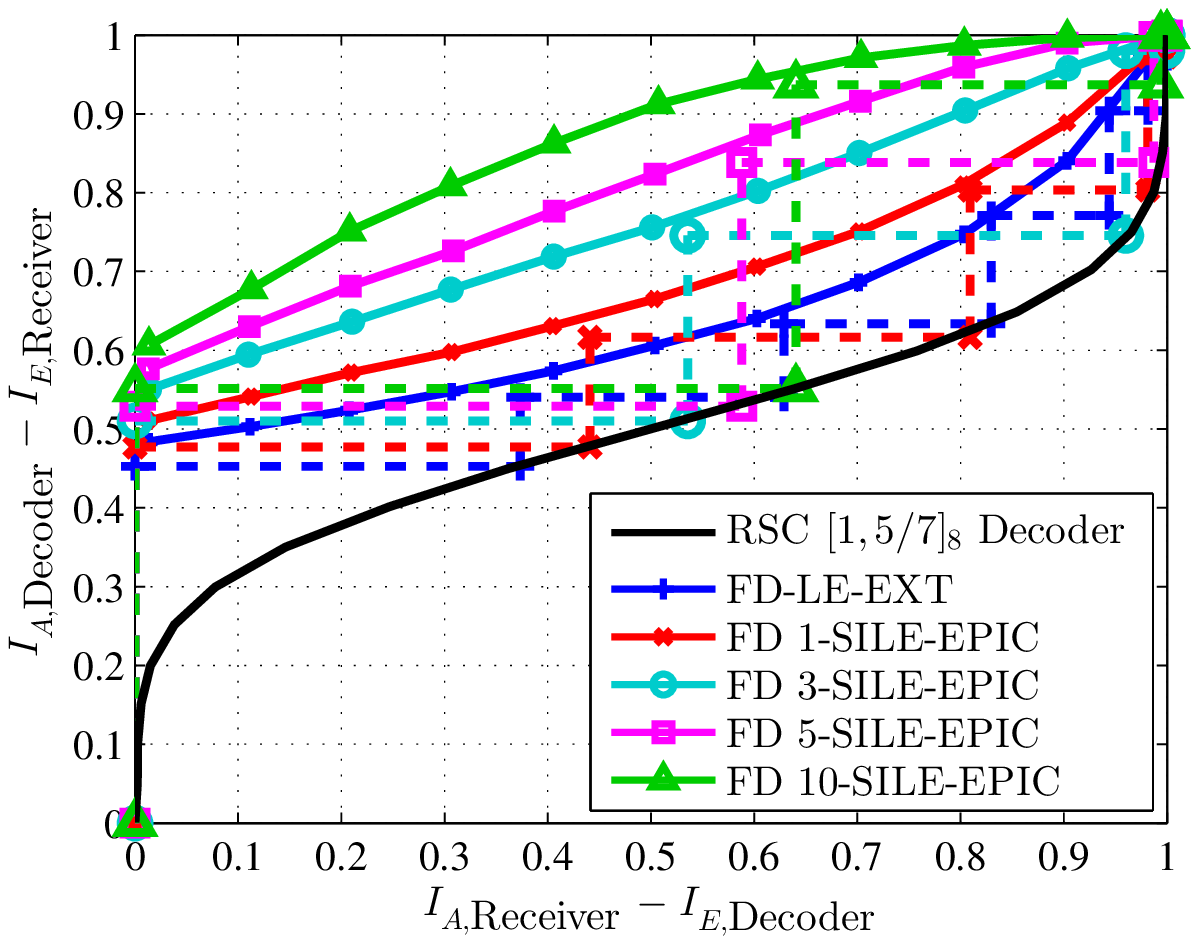}
    	\caption{EXIT curves and average MI trajectories of FDE with 8-PSK in Proakis C channel at $E_b/N_0=15$dB.}
    	\label{fig:qam16ProakisC_EXITtraj}
    \end{figure}
\fi

In Fig. \ref{fig:qam16ProakisC_EXITtraj}, EXIT charts of the proposed receiver, for $\mathcal{S}~=~0,1,3,5,10$, using a fixed linear damping
(see eq. (\ref{eq_demap_ep_filt_stats_new}))
, with $\beta~=~0.75$, is provided in solid curves, within the highly selective Proakis C channel, $\mathbf{h}~=~[1,2,3,2,1]/\sqrt{19}$, using the Gray-mapped 8-PSK constellation. Self-iterations are seen to significantly improve the MI for high $I_A$, which indicates a boosted convergence speed and an improved achievable rate. However, improvements for $I_A=0$ is relatively small, thus, the finite-length performance improvement will strongly depend on the EXIT chart of the decoder. In particular, for non-optimized but powerful turbo-like codes which have near-flat EXIT curves, improvement on the decoding threshold will be limited. However with a properly designed code, significant improvement would be possible.

This figure also shows the reverse transfer curve of the recursive systematic convolutional (RSC), code $[1, 5/7]_8$. Moreover, in dashed curves, the finite-length MI trajectories of this receiver with data blocks of length $K_b=768$ bits, using this channel decoder is plotted. The trajectories of the proposed EP-based receiver appears to follow the predicted transfer function fairly well, despite the short packet length, unlike the APP-based receivers as observed in \cite{sahinCipriano_2018_DFEEP}. This suggests that this receiver's EXIT analysis reflects its practical behaviour.

\ifdouble
    \begin{figure*}[t]
    	\centering
    	\includegraphics[width=7in]{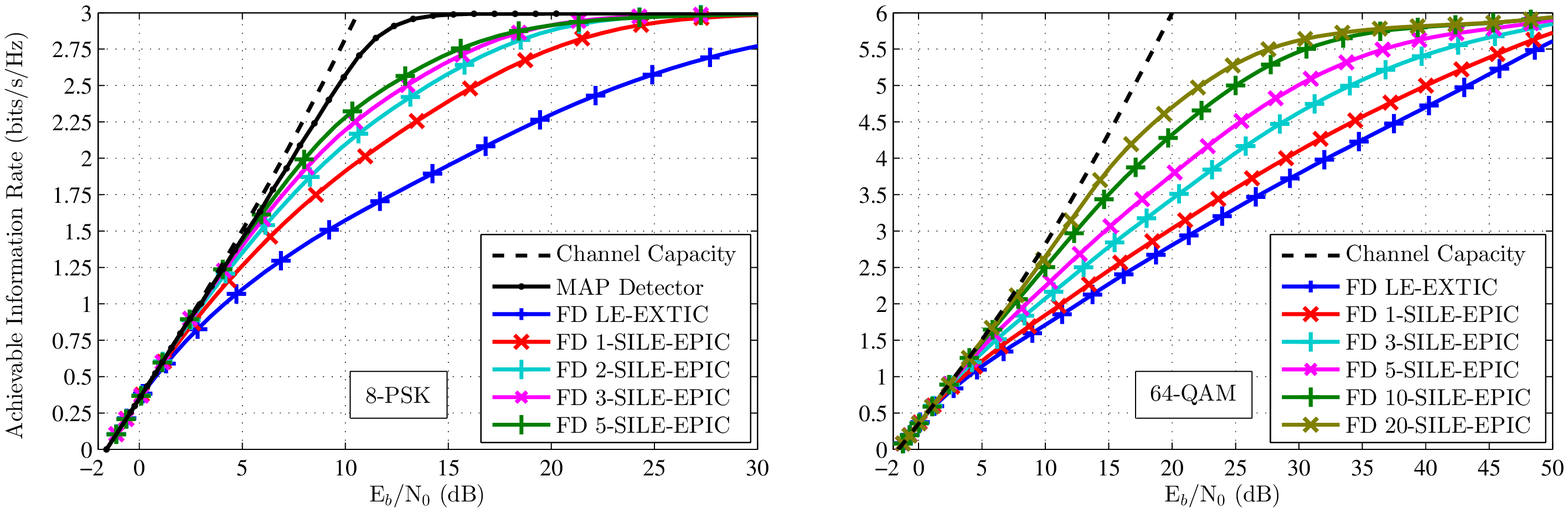}
	    \caption{Achievable rates of the proposed receiver in Proakis C with 8-PSK and 64-QAM.}
	    \label{fig_capa_proakisC}
    \end{figure*}
\else
    \begin{figure*}[t]
    	\centering
    	\includegraphics[width=6.5in]{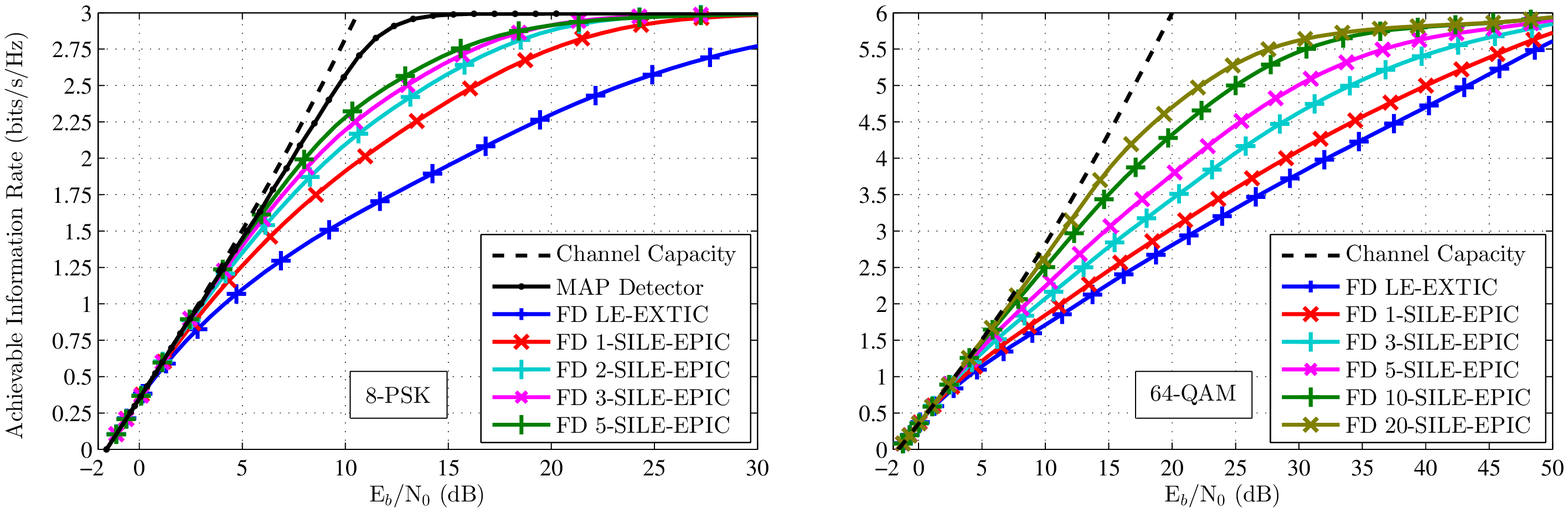}
	    \caption{Achievable rates of the proposed receiver in Proakis C with 8-PSK and 64-QAM.}
	    \label{fig_capa_proakisC}
    \end{figure*}
\fi

Another significant use of EXIT analysis is the evaluation of achievable rates (i.e. spectral efficiency) of the receiver. 
These values are numerically obtained using the area theorem of EXIT charts \cite{hagenauer_exit_2004}, and when considering the MAP detector, these rates constitute an accurate approximation of the channel symmetric information rate \cite{arnold_2006_simulation}, the highest possible transmission rate for practical constellations, without channel knowledge at the transmitter.
Achievable rates of the FD LE-EXTIC and the proposed receiver are given in Fig. \ref{fig_capa_proakisC}, for the Proakis C channel with 8-PSK and 64-QAM constellations. The Gaussian capacity of this channel, without transmit power optimization, is also plotted in dashed lines, it is computed using eq. (\ref{eq_chan_td}) with the vector-input AWGN channel capacity. 
Channel SIR with 8-PSK is given by the MAP detector curve in 8-PSK, but it is not plotted with 64-QAM due to the excessive computational resources it requires \cite{arnold_2006_simulation}. An exponential feature-based damping 
(see eq. (\ref{eq_demap_ep_damp_stats_new})) 
with $\beta=0.7\times 0.9^{s}$ is used for 8-PSK, whereas a fixed linear damping 
(see eq. (\ref{eq_demap_ep_filt_stats_new})) 
with $\beta=0.8$ is used for 64-QAM.

For 8-PSK, while the conventional FD LE-EXTIC \cite{tuchler_linear_2001}  follows the SIR limit within 0.5~dB up to $0.75$~bits/s/Hz, proposed EP-based self-iterations increase this range up to $2$~bits/s/Hz. In the 64-QAM case, FD LE-EXTIC follows the channel capacity within 1~dB up to $1$~bit/s/Hz and $3.33$~bits/s/Hz becomes achievable with 20 SI. 
For a rate-$1/2$ coded usage, the proposed receiver with $s\rightarrow +\infty$ brings over $3.9$~dB and $10.7$~dB improvement, over the conventional turbo FD LE in this channel, for respectively 8-PSK and 64-QAM constellations. These rates are achievable with properly designed coding schemes.

\subsection{Comparison with Single Tap FDE in Prior Work}

\ifdouble
    \begin{figure*}[t!]
    	\centering
    	\includegraphics[width=7in]{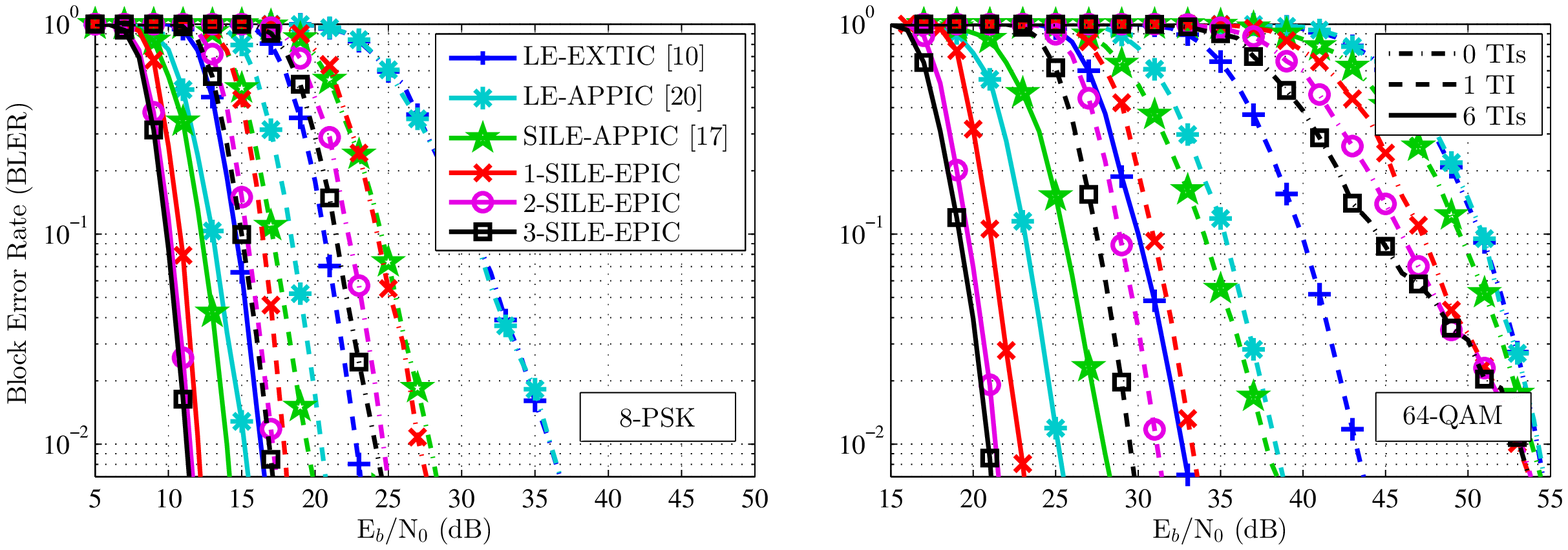}
    	\caption{BLER comparison of single-tap FD equalizers in Proakis C channel, with $K=256$ coded with rate-1/2 RSC $[1,5/7]_8$.
    	}
    	\label{fig_bler_proakisC}
    \end{figure*}
\else
    \begin{figure*}[t!]
    	\centering
    	\includegraphics[width=6.5in]{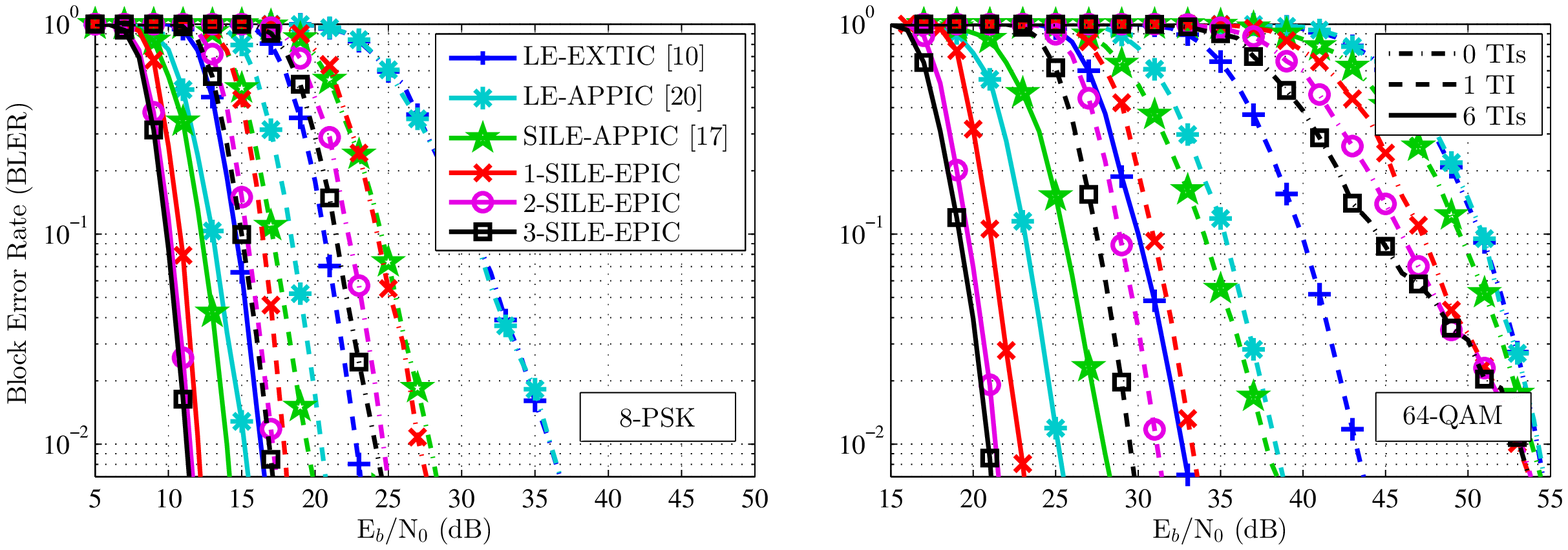}
    	\caption{BLER comparison of single-tap FD equalizers in Proakis C channel, with $K=256$ coded with rate-1/2 RSC $[1,5/7]_8$.
    	}
    	\label{fig_bler_proakisC}
    \end{figure*}
\fi

In this paragraph, observations in the previous section are completed with finite-length results within the same channel with a RSC code with soft MAP decoder. 
Block error rate (BLER) is obtained by Monte-Carlo simulations, with 30000 sent packets per point.
Unlike in asymptotic analysis, here we use dynamic damping that also depends on turbo-iterations, $\tau$, and accelerates convergence.
A feature-based damping with $\beta_{\tau,s}=0.7\times 0.9^{s+\tau}$ is used for 8-PSK, and a hybrid damping, consisting of a linear smoothing in the first TI, and feature-based damping afterwards, is applied with $\beta_{\tau,s}=0.85^{1+s+\tau}$, for 64-QAM.
Several single-tap FD equalizers are compared to our proposal in Fig.~\ref{fig_bler_proakisC}: the conventional linear equalizer \cite{tuchler_linear_2001} (\emph{LE-EXTIC}), the LE-IC with APP feedback \cite{visoz_frequency-domain_2006, witzke2002iterative} (\emph{LE-APPIC}), and the self-iterated LE-IC of \cite{tao_2015_singlecarrierfreq} (\emph{SILE-APPIC}).
The equalization complexity of these receivers is of the same order of computational complexity of $K\log_2 K$ at a given SI and TI, with slight differences underlying in the feedback computation.

Results show in Fig. \ref{fig_bler_proakisC} show that our proposal brings significant improvement on the decoding threshold, that grows with the number of SIs, at all TIs. On the contrary, multiple SIs with APP feedback degrades this threshold (not shown here due to lack of space). 
Without TI, 3~SIs bring respectively $9$~dB and $6$~dB gains for 8-PSK and 64-QAM, compared to LE-EXTIC, at $\text{BLER}=10^{-1}$. 
Performance in 64-QAM is limited at low PER without TIs, but our proposal with a single TI and 3~SIs reaches PER the prior work reach with 6~TIs, e.g. with six times lower decoding complexity.
Besides, asymptotically (6~TIs), SIs with EP bring over $8$~dB gain with respect to SILE-APPIC, and about $5$~dB gain over LE-APPIC, for 64-QAM, at $\text{BLER}=10^{-2}$. 
Compared to FD LE-EXTIC, 3~SIs bring around $4$~dB and $11.5$~dB gain, respectively for 8-PSK and 64-QAM, which is close to the 1/2-rate gains observed in the asymptotic analysis above.

These results encourage replacing TIs with SIs as demapping complexity is often insignificant relative to decoding.

\subsection{Comparison with EP-based Receivers in Prior Work}

\ifdouble
    \begin{figure*}[t!]
    	\centering
    	\includegraphics[width=7in]{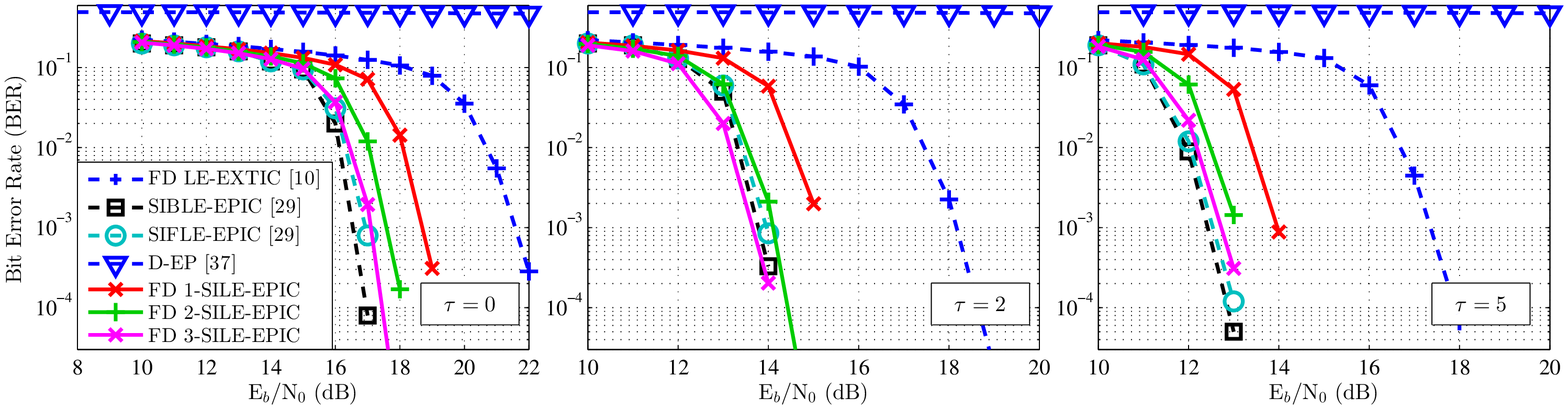}
    	\caption{BER comparison in Proakis C with 8-PSK, $K_d=4096$ and rate-1/2 regular $(3,6)$  LDPC code.
    	}
    	\label{fig_ber_ldpc}
    \end{figure*}
\else
    \begin{figure*}[t!]
    	\centering
    	\includegraphics[width=6.5in]{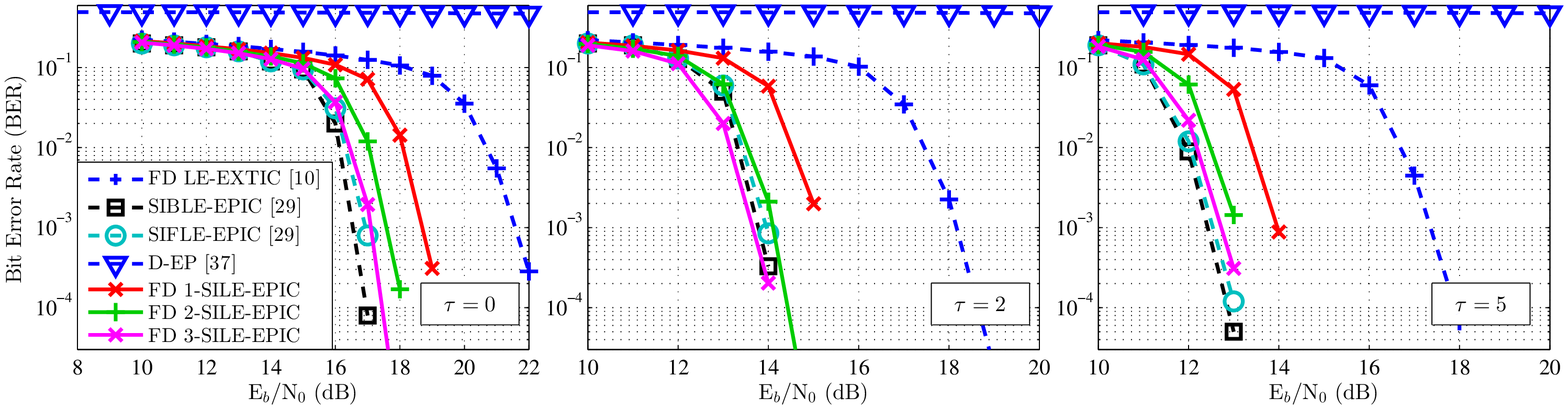}
    	\caption{BER comparison in Proakis C with 8-PSK, $K_d=4096$ and rate-1/2 regular $(3,6)$  LDPC code.
    	}
    	\label{fig_ber_ldpc}
    \end{figure*}
\fi

\ifdouble
    \begin{figure*}[t!]
	    \centering
	    \includegraphics[width=7in]{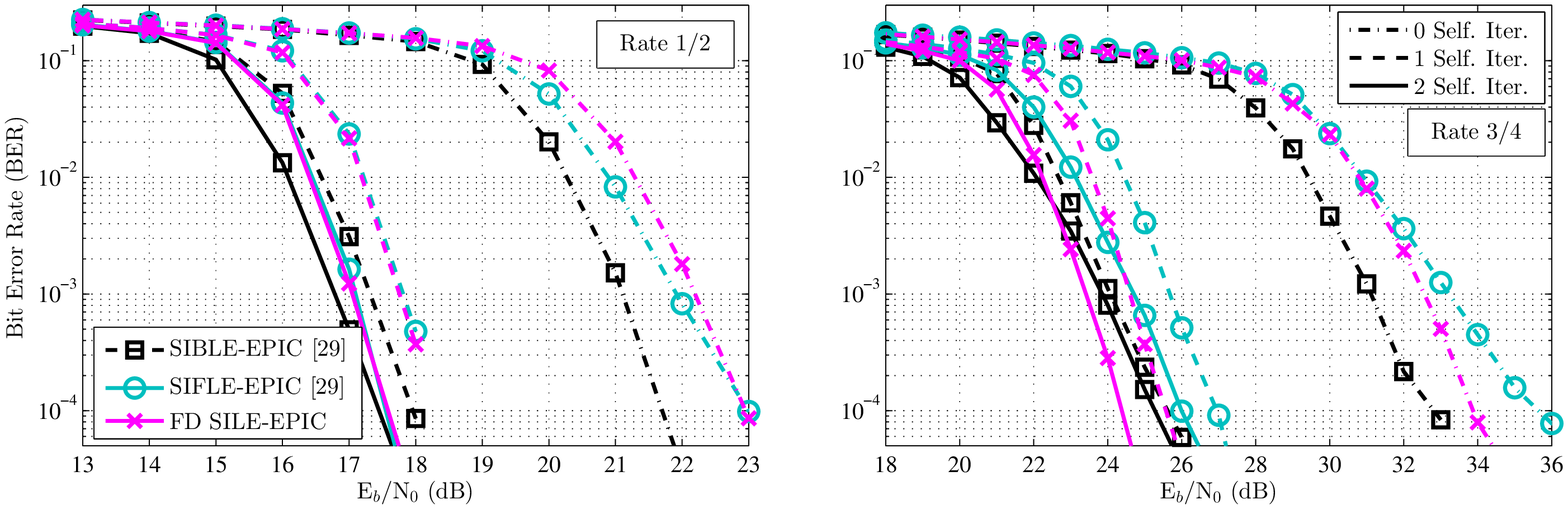}
	    \caption{SI LE-IC and FD SILE-IC in Proakis C with LDPC coded 16-QAM, with 5 turbo iterations.}
    	\label{fig:berLDPC16QAM_proakisC}
    \end{figure*}
\else
    \begin{figure*}[t!]
    	\centering
    	\includegraphics[width=6.5in]{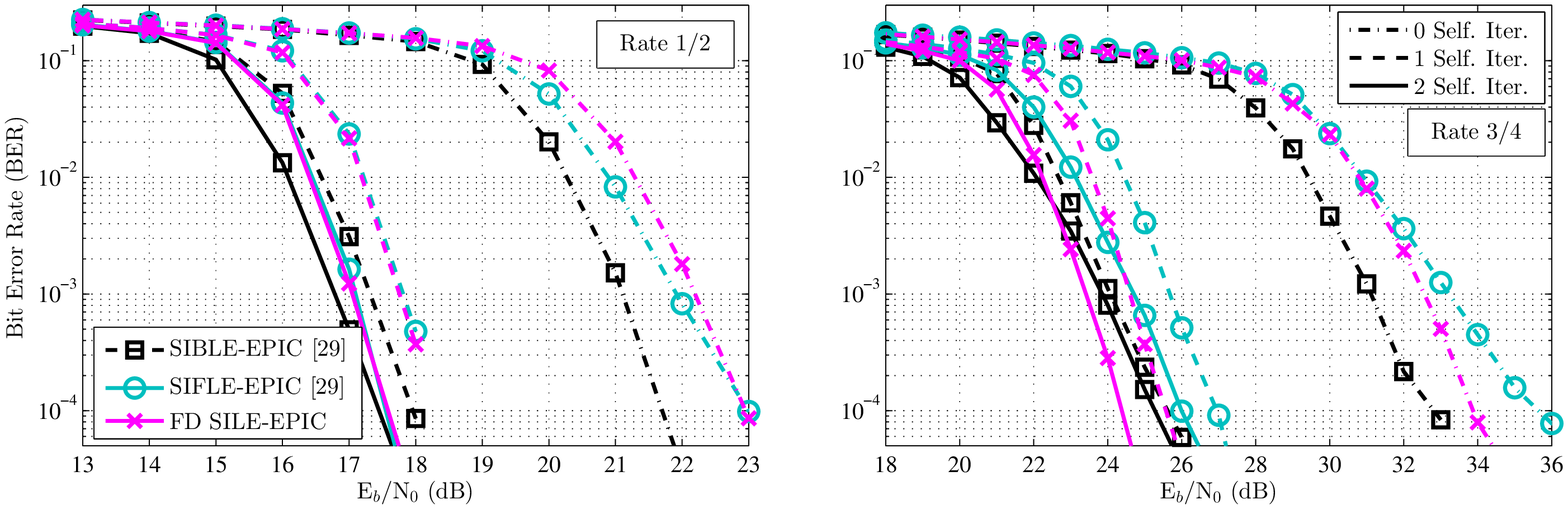}
    	\caption{SI LE-IC and FD SILE-IC in Proakis C with LDPC coded 16-QAM, with 5 turbo iterations.}
    	\label{fig:berLDPC16QAM_proakisC}
    \end{figure*}
\fi

\ifdouble
    \begin{figure}[t!]
    	\centering
    	\includegraphics[width=3.2in]{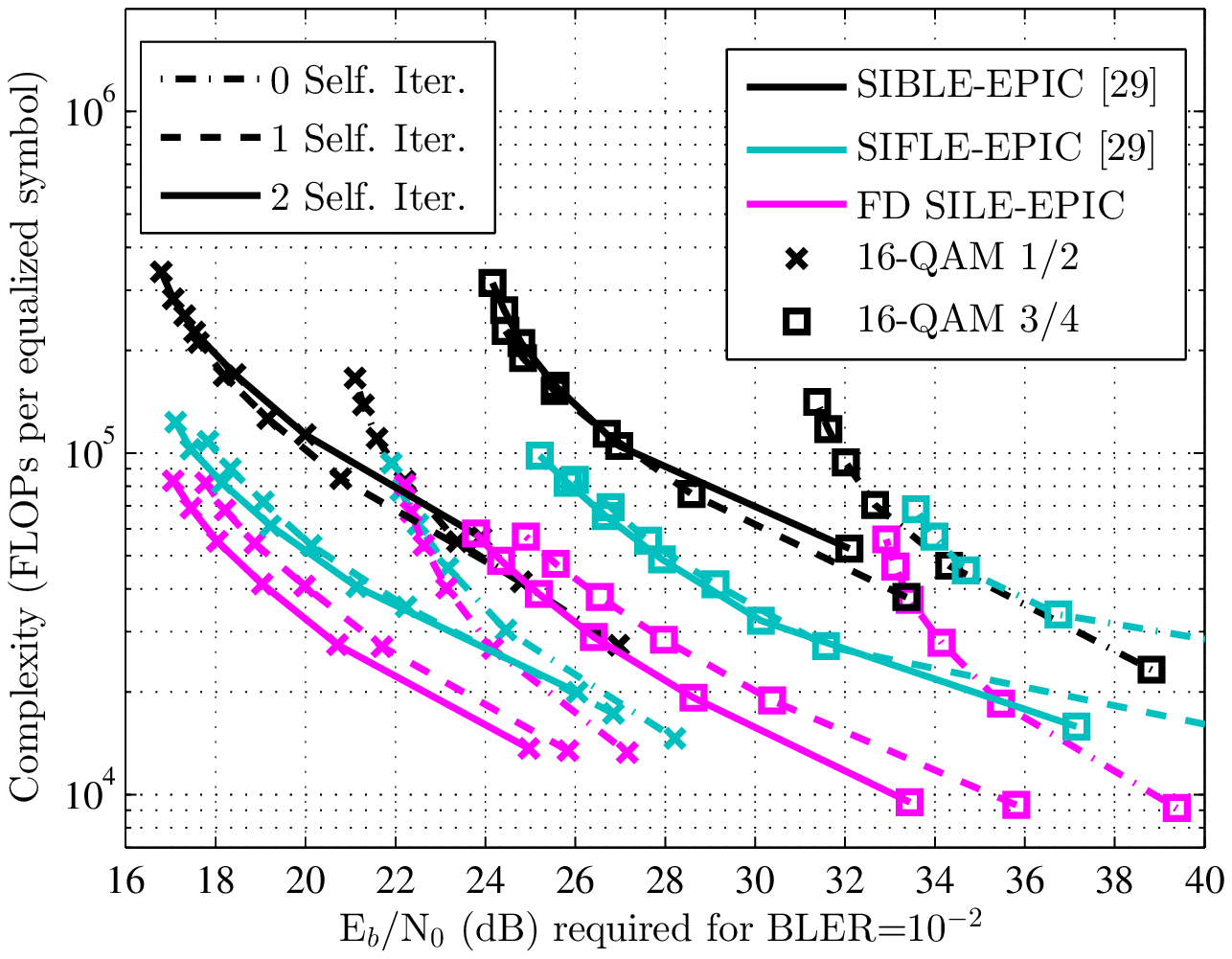}
    	\caption{Performance complexity trade-off for self-iterations in LDPC coded Proakis C.}
    	\label{fig:perfLDPCComplexityTradeoff_proakisC}
    \end{figure}
\else
    \begin{figure}[t!]
    	\centering
    	\includegraphics[width=3in]{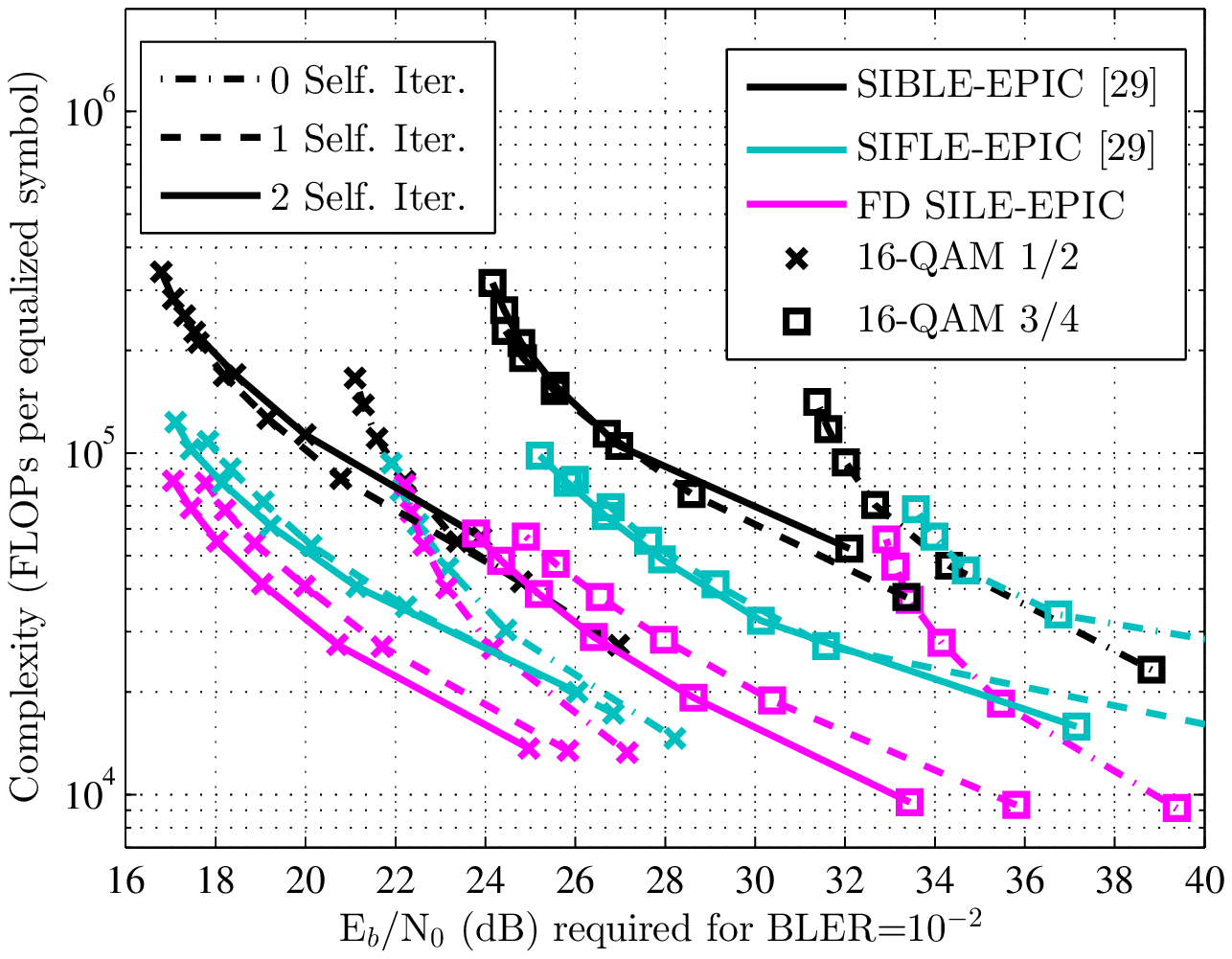}
    	\caption{Performance complexity trade-off for self-iterations in LDPC coded Proakis C.}
    	\label{fig:perfLDPCComplexityTradeoff_proakisC}
    \end{figure}
\fi

There are numerous emerging EP-based receivers in the literature, as stressed in the introduction, and in this section the proposed FDE is compared with self-iterated time-domain block (SIBLE-EPIC, denoted nuBEP in \cite{santosMurilloFuentes_2017_EPnuBLE}) and filter (SIFLE-EPIC, denoted EP-F in \cite{santosMurilloFuentes_2017_EPnuBLE}) receivers and to the single-tap FD receiver, D-EP, in \cite{wu_2017_spectralefficientbandallocation}. 
The proposed receiver is not compared to the exact FD receiver, J-EP in \cite{wu_2017_spectralefficientbandallocation, zhang_2016_expectationPropMIMOGFDM}, as it is equivalent to the SIBLE-EPIC with a single SI, without damping, making it sub-optimal compared to the SIBLE-EPIC. 
The block receiver in \cite{santosMurilloFuentes_2017_EPBLE}, is neither included in the comparison, as it is a sub-optimal block receiver which ignores prior information from the decoder at each SI (but a comparison is available in \cite{sahin_2018_IterativeEqEP_FDE}).

In Fig.~\ref{fig_ber_ldpc}, the bit error rate (BER) of the proposed receiver is compared with alternatives listed above. 
We consider 8-PSK constellation, and the low-density parity check (LDPC) coded Proakis C scenario from \cite{santosMurilloFuentes_2017_EPnuBLE}. 
The regular $(3,6)$ LDPC code is obtained by Progressive-Edge Growth (PEG) algorithm, and the decoder uses BP algorithm up to 100 iterations. The FD receiver, D-EP, cannot decode in Proakis-C channel, up to very high signal to noise ratios due to its sensitivity to channel nulls \cite[eq. (48)]{wu_2017_spectralefficientbandallocation}. 
Our FD proposal is seen to perform nearly as good as the TD EP-based receivers, with an order of computational complexity of $(\mathcal{S}+1)K\log_2K$ instead of $3LK^2$ (SIBLE-EPIC, 2~SIs) and or $27KL^2$ (SIFLE-EPIC, 2~SIs). For $\tau=5$, block and filter TD receivers have around $0.2$~dB gain over FD 3-SILE EPIC, but they are respectively around 500 and 16 times more complex.

Another LDPC-coded scenario in the Proakis C, with 16-QAM and with rate 1/2 and 3/4 encoding over $K_b~=~2048$ bits is reported in Fig. \ref{fig:berLDPC16QAM_proakisC}. All receiver use feature-based damping with the optimized parameter in \cite{santosMurilloFuentes_2017_EPnuBLE}, i.e. $\beta~=~ \min(0.3, 1-e^{\tau/1.5}/10)$. The regular $(3,12)$ LDPC code is also obtained by the PEG algorithm. In the rate-1/2 case, the proposed FDE is lower-bounded in BER by the block receiver, and following one SI, the difference between FD SILE-EPIC and SIFLE-EPIC is negligible. For the high rate case, at the right side of the figure, filter receiver's performance is over 1~dB worse for $\text{BER}<10^{-3}$, and although SIBLE-EPIC still has a better decoding threshold, it recovers less diversity than the proposed FD SILE-EPIC. This phenomenon should not be surprising, as exact receivers can be more prone to error propagation when decoder provides erroneous feedback, as also observed in filter receivers \cite{jeongMoon_2013_selfiteratingSoftEqualizer}.

These error rate results are completed with detailed computational complexity estimations in Fig. \ref{fig:perfLDPCComplexityTradeoff_proakisC}. 
This is evaluated with the number of multiply and accumulate units required to implement the receiver, estimated by the number of real additions and multiplications, amounting to half a floating point operation (0.5 FLOPs) each. 
Complexity is plotted versus the required bit SNR to decode transmitted blocks with $\text{BLER}=10^{-2}$, for $\tau=0,\dots,5$. These FLOP-counts also include the decoder complexity, which is considerably higher than equalizer complexity.
The proposed receiver performs overall efficient, both complexity and energy-wise, compared to the SIBLE-EPIC, with respectively  2.5, 4 and 5.4 times lower complexity for $\mathcal{S}=0, 1$ and 2 in the rate 3/4 case, and with respectively 2, 3.1 and 4.1 times lower complexity for the rate 1/2 case. This ratio is around ten times bigger, if the decoding complexity is not accounted for.

\begin{figure}[t!]
	\centering
	\includegraphics[width=3.2in]{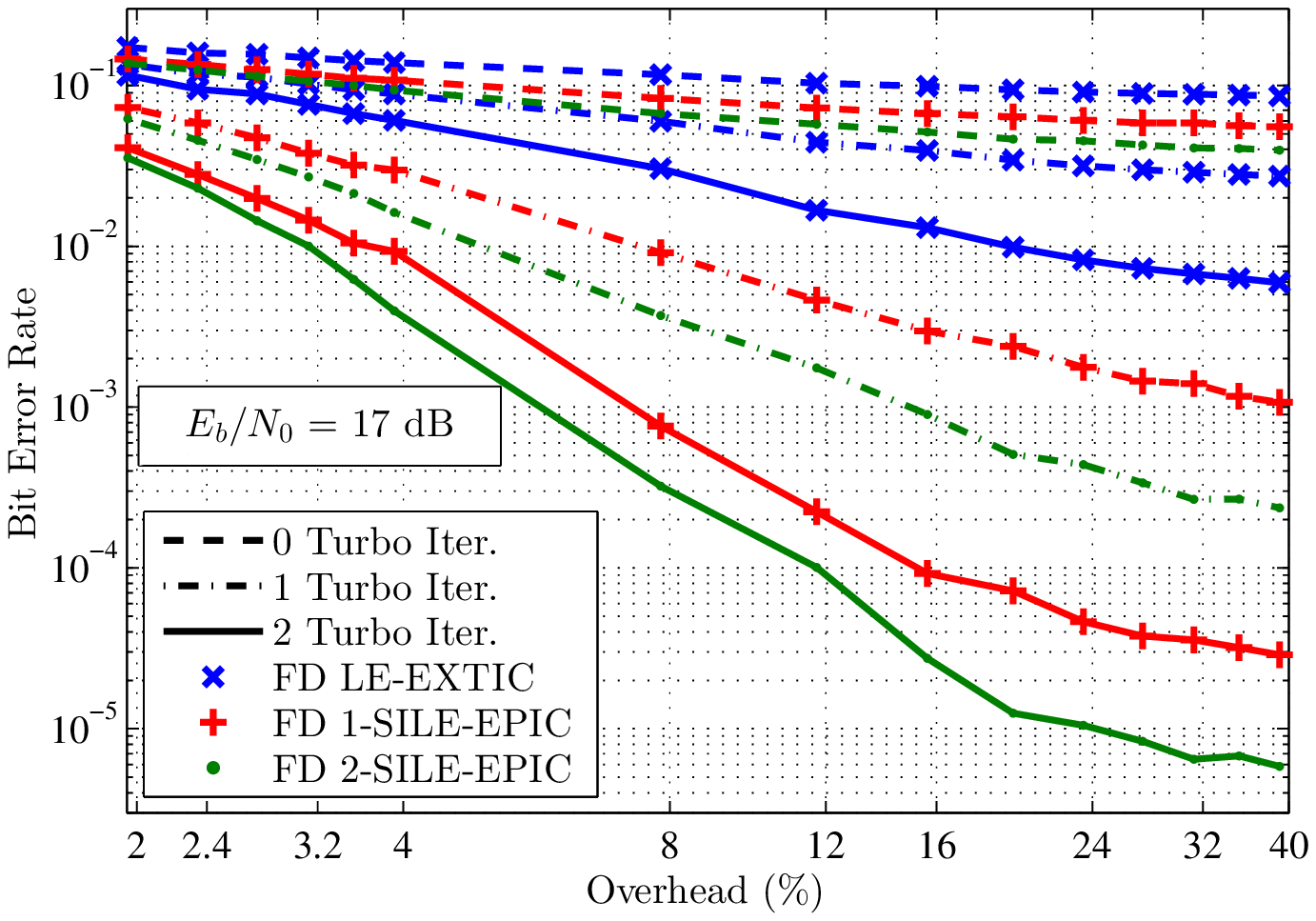}
	\caption{BER vs. overhead in Proakis C, with RSC $[1, 5/7]_8$ coded 8-PSK 
	}
	\label{fig_ber_imp_csir}
\end{figure}

\subsection{{On the Impact of Imperfect Channel Estimation}}

{
In this section, the performance of the proposed FD SILE-EPIC with imperfect channel estimates is evaluated. 
A mismatched receiver is considered to operate on an channel estimate $\mathbf{\hat{h}}$, corrupted by Gaussian noise $\pmb{\nu}$ whose variance $\sigma_\nu^2$ was selected using the model $\sigma_\nu^2 ={\sigma_w^2}/{(K_P \sigma_x^2)}$, where $K_P \geq L$ is the number of pilot symbols that would have been used for channel estimation in a complete receiver. 
We assume transmission of 8-PSK blocks with $K=256$ symbols and the quality of channel estimate is assessed via the overhead, defined by the ratio $K_P/K$.
}

{
Figure \ref{fig_ber_imp_csir} illustrates the behaviour of BER versus overhead, and the proposed EP-based receiver is shown to be robust to estimation errors.
Indeed, for a target BER of $10^{-2}$, a significant reduction of overhead is achieved with EP-based self-iterations; while baseline FDE with 2 TIs requires around 19\% overhead for channel estimation, using our proposal, one turbo and one self iteration requires only 8\% overhead and one self and 2 TIs requires 4\%.
Thus PHY data frames with shorter number of pilot symbols could be designed to increase spectral efficiency.
}

\subsection{{Comparison with work on Approximate Message Passing}}

{
AMP is a commonly used technique in signal processing fields such as compressed sensing or data classification, which is based on belief propagation, often with Gaussian approximation for tractable MMSE estimation. 
In particular, Generalized AMP (GAMP) is adapted for linear probabilistic models as in Eq. (\ref{eq_circmodel}) \cite{rangan_2010_GAMP}, however it is designed for fixed priors and cannot be directly applied in the context of turbo detection. In \cite{guo_2013_iterativeFDE_GAMPext}, Guo \emph{et. al.} reworked GAMP for turbo-equalization by generating extrinsic outputs (GAMP-ext), and derived a self-iterated FDE. 

Considering that GAMP algorithm is derived using loopy belief propagation, a recent improved extension based on EP is Vector AMP (VAMP) \cite{rangan_2016_VAMP}.
In order to draw parallel's between AMP-based algorithms and the proposed EP-based framework, similarly to \cite{guo_2013_iterativeFDE_GAMPext}, VAMP algorithm can be tweaked to operate with extrinsic outputs, and extended to circularly complex Gaussian distributions to be applied on frequency domain observations (VAMP-ext).

\begin{figure*}[t!]
	\centering
	\includegraphics[width=6in]{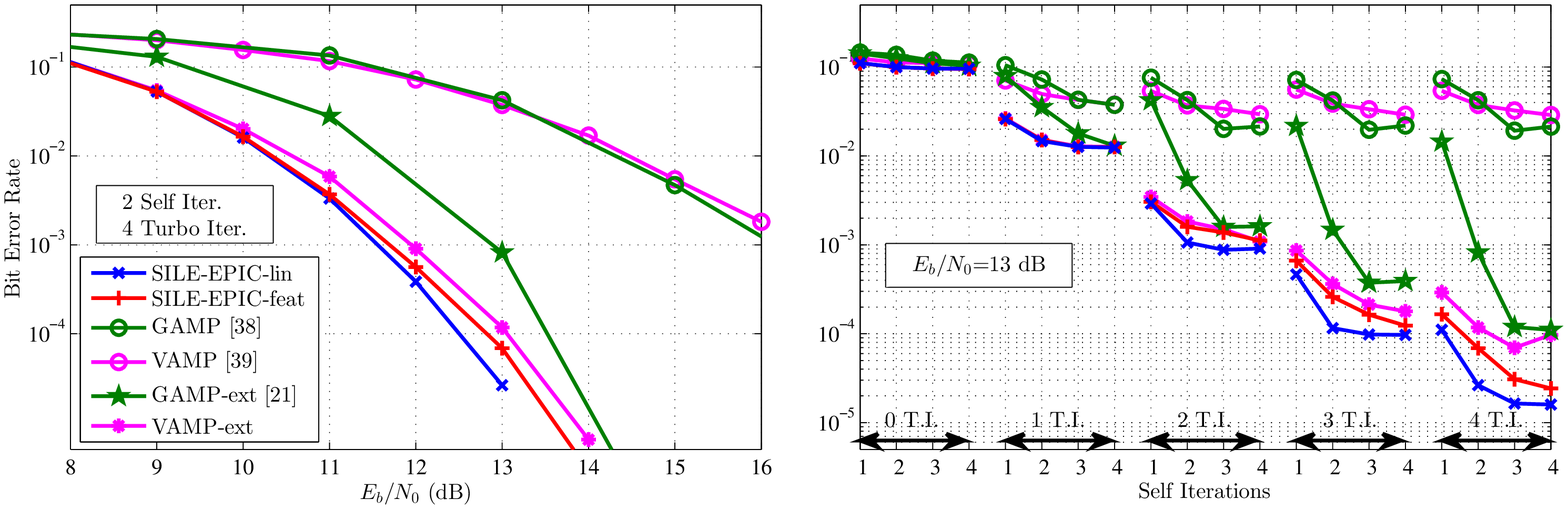}
	\caption{Comparison of iterative receivers based on AMP with the proposed FD SILE EPIC in Proakis C, with RSC $[1, 5/7]_8$ coded 8-PSK and with static BER-optimized damping for each receiver, at each iteration and SNR}
	\label{fig_ber_fde_comp_amp}
\end{figure*}

Derivation details of VAMP-ext are not given due to lack of space, and it results in a similar algorithm to FD SILE-EPIC, but with notable differences in damping and convergence heuristics.
In Figure \ref{fig_ber_fde_comp_amp}, FDEs based on AMP are compared with the proposed FD SILE-EPIC with both linear (``-lin", see eq. (\ref{eq_demap_ep_filt_stats_new})) and feature-based damping (``-feat", see eq. (\ref{eq_demap_ep_damp_stats_new})).
Numerical results indicate that our original proposal FD SILE-EPIC converges to further lower error rates than AMP-based alternative, with over 1~dB gain on GAMP-ext, and over 0.6~dB on VAMP-ext.
These results show that AMP-based algorithms themselves are not well-adapted to turbo-detection use-case, and that it is preferable to address such systems using the founding theory of EP. 
}

\subsection{Conclusion on EP-based SC-FDE}

Finite-length error rate performance and the asymptotic analysis show that the proposed SC-FDE receiver, obtained by the considered EP-based message passing framework in section \ref{sec:fdsile}, outperforms similar alternative receivers (single-tap FDE), and performs almost identically to the exact TD receivers while having a significantly lower computational complexity.

EP with multivariate white Gaussian distributions is exposed in this elementary SC-FDE system, to improve readability and to simplify performance analysis. In the following, we give an overview of application of this framework to more complex communication systems.

\section{Application to Time-Varying Channel Equalization: Overlap FDE}\label{sec:overlapFDE}

\subsection{System Model}

A notable issue of FDE is its inability to mitigate time-varying channels whose coherence time is shorter than the processing block duration. In this case the FD channel matrix is no longer diagonal, and inter-carrier interference is generated.
\emph{Overlap FDE} is a possible approach for mitigating problems above, without significantly increasing the receiver complexity.

This technique consists in using $N$-point FFTs, with $N < K$, to carry out baseband processing, on \emph{virtual} overlapping sub-blocks of received samples \cite{vollmerHaardtGotze_2001_comparativeDetectionTDCDMA,martoyoWeiss_2003_LowComplexityCDMA}. This strategy inherently generates IBI between sub-blocks, which is mitigated either by selecting an appropriate sub-block length $N$, or by using additional signal processing. Some recent usage examples include its usage with faster-than-Nyquist signalling \cite{fukumoto_2015_overlap}, and with  doubly selective channels \cite{ravaeiNejad_2017_LowComplexyfFDTEQHF}. 
In this section, various EP-based overlap FDE receivers are derived and evaluated.

\subsection{Conventional Overlap FDE with ``No-Interference"}

\ifdouble
    \begin{figure}[!t]
    	\centering
    	\rotatebox{90}{\begin{tikzpicture}
    	[
    	xscale = 0.95,% to scale horizontally everything but the text
    	yscale = 1.4, every node/.style={xscale=1.05, yscale=1.05}
    	]
    
    	% NODES DEFINITION				
    	\node (fr_ul)[coordinate] at (-2.25,1.5){};
    	\node (fr_ur)[coordinate] at (1.75,1.5){};
    	\node (fr_dr)[coordinate] at (1.75,1){};
    	\node (fr_dl)[coordinate] at (-2.25,1){};
    	\node (fr_label) at (-0.25,1.25){received block $\mathbf{{y}}$};
    	
    	\node (sbl1_ul)[coordinate] at (-2,0.6){};
    	\node (sbl1_ur)[coordinate] at (0,0.6){};
    	\node (sbl1_dr)[coordinate] at (0,0.1){};
    	\node (sbl1_dl)[coordinate] at (-2,0.1){};
    
    	\node (sble1_ul)[coordinate] at (-2,-0.3){};
    	\node (sble1_ur)[coordinate] at (0,-0.3){};
    	\node (sble1_dr)[coordinate] at (0,-0.8){};
    	\node (sble1_dl)[coordinate] at (-2,-0.8){};

    	\node (sbl2_ul)[coordinate] at (-0.5,-1.5){};
    	\node (sbl2_ur)[coordinate] at (1.5,-1.5){};
    	\node (sbl2_dr)[coordinate] at (1.5,-2){};
    	\node (sbl2_dl)[coordinate] at (-0.5,-2){};
    	
    	\node (sble2_ul)[coordinate] at (-0.5,-2.25){};
    	\node (sble2_ur)[coordinate] at (1.5,-2.25){};
    	\node (sble2_dr)[coordinate] at (1.5,-2.75){};
    	\node (sble2_dl)[coordinate] at (-0.5,-2.75){};
    	
    	\node (fr2_ul)[coordinate] at (-2.25,-3){};
    	\node (fr2_ur)[coordinate] at (1.75,-3){};
    	\node (fr2_dr)[coordinate] at (1.75,-3.5){};
    	\node (fr2_dl)[coordinate] at (-2.25,-3.5){};

    	% PATHS
    	\draw [-,dashed] (sbl1_ul) -- (-2,1);
    	\draw [-,dashed] (sbl1_ur) -- (0,1);
        \draw [-,dotted] (sbl2_ul) -- (-0.5,1);
    	\draw [-,dotted] (sbl2_ur) -- (1.5,1);
        
    	\draw [-,thick] (fr_ul) -- (fr_ur);
    	\draw [-,thick] (fr_dr) -- (fr_dl);
    	\draw [dashed] plot [smooth, tension=0.75] coordinates { (-2.25,1.5) (-2.3,1.375) (-2.2,1.125) (-2.25,1)};
        \draw [dashed] plot [smooth, tension=0.75] coordinates { (1.75,1.5) (1.7,1.375) (1.8,1.125) (1.75,1)};
    
    	\draw [<->] (-2,0.7) -- node[pos=0.5,yshift=0.175cm]{\scriptsize$n^\text{{th}}$ sub-block}(0,0.7);
    	\draw [-,thick,fill=white] (sbl1_ul) -- (sbl1_ur) -- (sbl1_dr) -- (sbl1_dl) -- (sbl1_ul);
    	\node (sbl1_label) at (-1,0.4){\footnotesize$\mathbf{{\tilde{y}}}_{(n-1)N_d}$};
    	
    	\draw [->, thick,double] (-1,0.1) -- node[xshift=1cm]{\footnotesize $N$-point FDE}(-1,-0.3);
        \draw [-,thick,fill=white] (sble1_ul) -- (sble1_ur) -- (sble1_dr) -- (sble1_dl) -- (sble1_ul);
    	\node (sble1_label) at (-1.05,-0.55){\footnotesize$\mathbf{{\tilde{\hat{x}}}}_{(n-1)N_d}$};
    	
    	\draw [-,densely dashed] (-1.8,-0.3) -- (-1.8,-3.5);
    	\draw [-,densely dashed] (-0.31,-0.3) -- (-0.31,-3.5);
    
    	\draw [<->] (-0.5,-1.4) -- node[pos=0.65,yshift=0.35cm]{\parbox[c]{1.2cm}{\scriptsize$(n+1)^\text{{th}}$\\ sub-block}}(1.5,-1.4);
    	\draw [-,thick,fill=white] (sbl2_ul) -- (sbl2_ur) -- (sbl2_dr) -- (sbl2_dl) -- (sbl2_ul);
    	\node (sbl1_label) at (0.5,-1.7){\footnotesize$\mathbf{{\tilde{y}}}_{nN_d}$};
    	\draw [->, thick,double] (0.5,-2) -- (0.5,-2.25);
    	\draw [-,thick,fill=white] (sble2_ul) -- (sble2_ur) -- (sble2_dr) -- (sble2_dl) -- (sble2_ul);
        \node (sble_label) at (0.5,-2.5){\footnotesize$\mathbf{{\tilde{\hat{x}}}}_{nN_d}$};
    	
        \draw [-,densely dotted] (-0.29,-2.25) -- (-0.29,-3.5);
    	\draw [-,densely dotted] (1.2,-2.25) -- (1.2,-3.5);

    	\draw [->,double] (-1,-0.8) -- node[xshift=-0.24cm, rotate=270]{\scriptsize Extract $N_d$ symbols}(-1,-3);
    	\draw [->,double] (0.5,-2.75) -- (0.5,-3);

        \draw [white,fill=white] (fr2_ul) -- (fr2_ur) -- (fr2_dr) -- (fr2_dl) -- (fr2_ul);
        \draw [-,thick] (fr2_ul) -- (fr2_ur);
    	\draw [-,thick] (fr2_dr) -- (fr2_dl);
    	\draw [dashed] plot [smooth, tension=0.75] coordinates { (-2.25,-3.5) (-2.3,-3.375) (-2.2,-3.125) (-2.25,-3)};
        \draw [dashed] plot [smooth, tension=0.75] coordinates { (1.75,-3.5) (1.7,-3.375) (1.8,-3.125) (1.75,-3)};
    	\node (fr2_label) at (-0.25,-3.25){equalized block $\mathbf{{{\hat{x}}}}$};
    
    	% AUX
    	\begin{scope}[]
        \draw [dashed, fill=gray, fill opacity=0.1] (sble1_dl) |- (-1.8,-0.3) |- (sble1_dl){};
    	\draw [dashed, fill=gray, fill opacity=0.1] (sble1_dr) |- (-0.31,-0.3) |- (sble1_dr){};
        \draw [dashed, fill=gray, fill opacity=0.1] (sble2_dl) |- (-0.29,-2.25) |- (sble2_dl){};
    	\draw [dashed, fill=gray, fill opacity=0.1] (sble2_dr) |- (1.2,-2.25) |- (sble2_dr){};
    	\end{scope}
    
        \draw [<->] (-2,-0.9) -- node[pos=0.3, yshift=-0.2cm, rotate=270]{\scriptsize $N_l$}(-1.8,-0.9) {};
        \draw [<->] (-0.31,-0.9) -- node[pos=0.5, yshift=-0.2cm, rotate=270]{\scriptsize $N_r$}(0,-0.9) {};
    
    	\end{tikzpicture}}
    	\caption{Overlap FDE processing scheme with sub-blocks.}
    	\label{fig_tr}
    \end{figure}
\else
    \begin{figure}[!t]
    	\centering
    	\rotatebox{90}{\begin{tikzpicture}
    	[
    	xscale = 0.95,% to scale horizontally everything but the text
    	yscale = 1.4, every node/.style={xscale=1.05, yscale=1.05}
    	]
    
    	% NODES DEFINITION				
    	\node (fr_ul)[coordinate] at (-2.25,1.5){};
    	\node (fr_ur)[coordinate] at (1.75,1.5){};
    	\node (fr_dr)[coordinate] at (1.75,1){};
    	\node (fr_dl)[coordinate] at (-2.25,1){};
    	\node (fr_label) at (-0.25,1.25){received block $\mathbf{{y}}$};
    	
    	\node (sbl1_ul)[coordinate] at (-2,0.6){};
    	\node (sbl1_ur)[coordinate] at (0,0.6){};
    	\node (sbl1_dr)[coordinate] at (0,0.1){};
    	\node (sbl1_dl)[coordinate] at (-2,0.1){};
    
    	\node (sble1_ul)[coordinate] at (-2,-0.3){};
    	\node (sble1_ur)[coordinate] at (0,-0.3){};
    	\node (sble1_dr)[coordinate] at (0,-0.8){};
    	\node (sble1_dl)[coordinate] at (-2,-0.8){};

    	\node (sbl2_ul)[coordinate] at (-0.5,-1.5){};
    	\node (sbl2_ur)[coordinate] at (1.5,-1.5){};
    	\node (sbl2_dr)[coordinate] at (1.5,-2){};
    	\node (sbl2_dl)[coordinate] at (-0.5,-2){};
    	
    	\node (sble2_ul)[coordinate] at (-0.5,-2.25){};
    	\node (sble2_ur)[coordinate] at (1.5,-2.25){};
    	\node (sble2_dr)[coordinate] at (1.5,-2.75){};
    	\node (sble2_dl)[coordinate] at (-0.5,-2.75){};
    	
    	\node (fr2_ul)[coordinate] at (-2.25,-3){};
    	\node (fr2_ur)[coordinate] at (1.75,-3){};
    	\node (fr2_dr)[coordinate] at (1.75,-3.5){};
    	\node (fr2_dl)[coordinate] at (-2.25,-3.5){};

    	% PATHS
    	\draw [-,dashed] (sbl1_ul) -- (-2,1);
    	\draw [-,dashed] (sbl1_ur) -- (0,1);
        \draw [-,dotted] (sbl2_ul) -- (-0.5,1);
    	\draw [-,dotted] (sbl2_ur) -- (1.5,1);
        
    	\draw [-,thick] (fr_ul) -- (fr_ur);
    	\draw [-,thick] (fr_dr) -- (fr_dl);
    	\draw [dashed] plot [smooth, tension=0.75] coordinates { (-2.25,1.5) (-2.3,1.375) (-2.2,1.125) (-2.25,1)};
        \draw [dashed] plot [smooth, tension=0.75] coordinates { (1.75,1.5) (1.7,1.375) (1.8,1.125) (1.75,1)};
    
    	\draw [<->] (-2,0.7) -- node[pos=0.5,yshift=0.175cm]{\scriptsize$n^\text{{th}}$ sub-block}(0,0.7);
    	\draw [-,thick,fill=white] (sbl1_ul) -- (sbl1_ur) -- (sbl1_dr) -- (sbl1_dl) -- (sbl1_ul);
    	\node (sbl1_label) at (-1,0.4){\footnotesize$\mathbf{{\tilde{y}}}_{(n-1)N_d}$};
    	
    	\draw [->, thick,double] (-1,0.1) -- node[xshift=1cm]{\footnotesize $N$-point FDE}(-1,-0.3);
        \draw [-,thick,fill=white] (sble1_ul) -- (sble1_ur) -- (sble1_dr) -- (sble1_dl) -- (sble1_ul);
    	\node (sble1_label) at (-1.05,-0.55){\footnotesize$\mathbf{{\tilde{\hat{x}}}}_{(n-1)N_d}$};
    	
    	\draw [-,densely dashed] (-1.8,-0.3) -- (-1.8,-3.5);
    	\draw [-,densely dashed] (-0.31,-0.3) -- (-0.31,-3.5);
    
    	\draw [<->] (-0.5,-1.4) -- node[pos=0.65,yshift=0.45cm]{\parbox[c]{1.2cm}{\scriptsize$(n+1)^\text{{th}}$\\ sub-block}}(1.5,-1.4);
    	\draw [-,thick,fill=white] (sbl2_ul) -- (sbl2_ur) -- (sbl2_dr) -- (sbl2_dl) -- (sbl2_ul);
    	\node (sbl1_label) at (0.5,-1.7){\footnotesize$\mathbf{{\tilde{y}}}_{nN_d}$};
    	\draw [->, thick,double] (0.5,-2) -- (0.5,-2.25);
    	\draw [-,thick,fill=white] (sble2_ul) -- (sble2_ur) -- (sble2_dr) -- (sble2_dl) -- (sble2_ul);
        \node (sble_label) at (0.5,-2.5){\footnotesize$\mathbf{{\tilde{\hat{x}}}}_{nN_d}$};
    	
        \draw [-,densely dotted] (-0.29,-2.25) -- (-0.29,-3.5);
    	\draw [-,densely dotted] (1.2,-2.25) -- (1.2,-3.5);

    	\draw [->,double] (-1,-0.8) -- node[xshift=-0.24cm, rotate=270]{\scriptsize Extract $N_d$ symbols}(-1,-3);
    	\draw [->,double] (0.5,-2.75) -- (0.5,-3);

        \draw [white,fill=white] (fr2_ul) -- (fr2_ur) -- (fr2_dr) -- (fr2_dl) -- (fr2_ul);
        \draw [-,thick] (fr2_ul) -- (fr2_ur);
    	\draw [-,thick] (fr2_dr) -- (fr2_dl);
    	\draw [dashed] plot [smooth, tension=0.75] coordinates { (-2.25,-3.5) (-2.3,-3.375) (-2.2,-3.125) (-2.25,-3)};
        \draw [dashed] plot [smooth, tension=0.75] coordinates { (1.75,-3.5) (1.7,-3.375) (1.8,-3.125) (1.75,-3)};
    	\node (fr2_label) at (-0.25,-3.25){equalized block $\mathbf{{{\hat{x}}}}$};
    
    	% AUX
    	\begin{scope}[]
        \draw [dashed, fill=gray, fill opacity=0.1] (sble1_dl) |- (-1.8,-0.3) |- (sble1_dl){};
    	\draw [dashed, fill=gray, fill opacity=0.1] (sble1_dr) |- (-0.31,-0.3) |- (sble1_dr){};
        \draw [dashed, fill=gray, fill opacity=0.1] (sble2_dl) |- (-0.29,-2.25) |- (sble2_dl){};
    	\draw [dashed, fill=gray, fill opacity=0.1] (sble2_dr) |- (1.2,-2.25) |- (sble2_dr){};
    	\end{scope}
    
        \draw [<->] (-2,-0.9) -- node[pos=0.3, yshift=-0.2cm, rotate=270]{\scriptsize $N_l$}(-1.8,-0.9) {};
        \draw [<->] (-0.31,-0.9) -- node[pos=0.5, yshift=-0.2cm, rotate=270]{\scriptsize $N_r$}(0,-0.9) {};
    
    	\end{tikzpicture}}
    	\caption{Overlap FDE processing scheme with sub-blocks.}
    	\label{fig_tr}
    \end{figure}
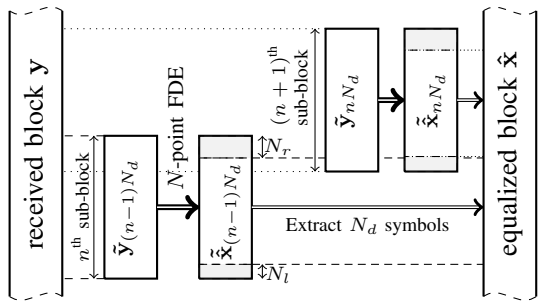
\fi

Overlap FDE, also called FDE with overlap-and-save or overlap-and-cut, carries out a linear deconvolution with multiple circular convolutions.
Given a signal block $\mathbf{v}\in\mathbb{C}^K$, its $N$-point sub-blocks are denoted $\tilde{\mathbf{v}}_{k}=[v_{k},\dots,v_{k+N-1}]^T$, with $v_{k}=0$, for all $k<0$ or $k\geq K$. SC-FDE model with sub-blocks is written as
\begin{equation}
\tilde{\mathbf{y}}_{k} = \mathbf{H}_k\tilde{\mathbf{x}}_{k} + \mathbf{G}_k(\tilde{\mathbf{x}}_{k-N}-\tilde{\mathbf{x}}_{k}) + \tilde{\mathbf{w}}_{k},
\end{equation}
where $\mathbf{H}_k$ is a $N \times N$ circular channel matrix as in eq. (\ref{eq_circmodel}), and $\mathbf{G}_k$ is an $N \times N$ matrix, whose $L-1$ upper diagonals are equal to those of $\mathbf{H}_k$, and other elements are zeros. Unlike the channel model in eq. (\ref{eq_circmodel}), here the channel may quasi-statically vary between sub-blocks. Hence with a small enough $N$, a time-varying frequency selective channel can be approximated by this model.

%$N$-point 
SC-FDE is used on sub-blocks, by \emph{ignoring the IBI term}, and $N_l$ symbols from the head and $N_r$ symbols from the tail of the equalized sub-block are thrown away. 
$N_l+N_r$ symbols are overlapping between two successive sub-blocks, as shown in Fig.~\ref{fig_tr}, and by extracting the remaining $N_d=N-N_l-N_r$ symbols, this procedure is repeated for $N_b=\lceil K/N_d \rceil$ sub-blocks in parallel. For extending this scheme to use the proposed EP-based framework from the previous sections; one could implement each $N_b$ equalizer of length $N$ using FD SILE-EPIC. Hence each sub-equalizer would have its own self-iteration loops, and independently evolving estimate variances. 
But as BICM is used across all sub-blocks, differences of estimate variances between sub-blocks is small, hence, for simplicity, all the sub-equalizers (FD SILE-EPIC) are assumed to use a common SI loop, with the common output variance denoted $v^e \triangleq N_b^{-1} \sum_{n=1}^{N_b}{v}^e_n$, where $v^e_n$ is the $n^\text{th}$ sub-equalizer's output variance, and the common feedback variance is $v^d$.

We denote this overlap FDE scheme, no-interference (NI), its performance at a given signal-to-noise ratio (SNR) can be close to that of SC-FDE with $K$-point FFT, if $N$, $N_l$ and $N_r$ are sufficient to remove all residual IBI over $N_d$ extracted symbols \cite{vollmerHaardtGotze_2001_comparativeDetectionTDCDMA,martoyoWeiss_2003_LowComplexityCDMA,fukumoto_2015_overlap}. However, for moderately or highly selective channels, the IBI spread can be very large, requiring $N\gg N_d$. Otherwise, residual IBI is present, and causes detection errors, whose occurrence increase with the SNR, due to interference enhancement caused by mismatched filter weights.

\ifdouble
    \begin{figure*}[t]
    	\centering
    	\includegraphics[width=7in]{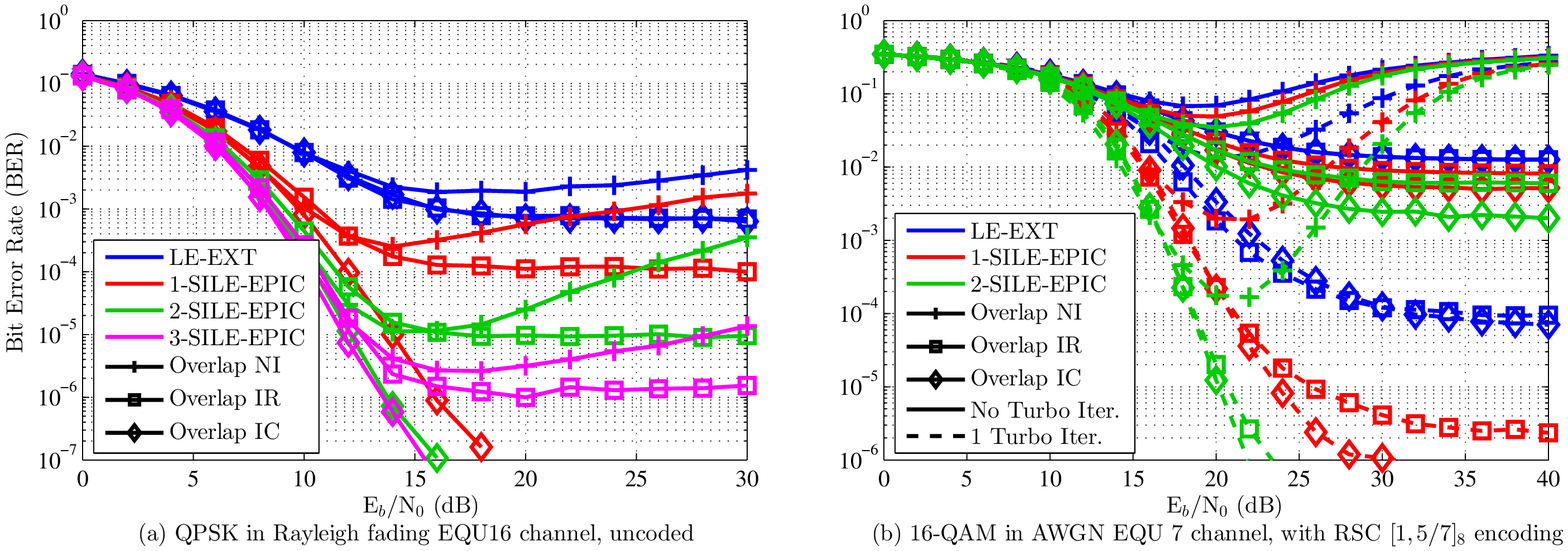}
    	\caption{IBI mitigation capabilities of overlap FDE schemes.}
    	\label{fig_overlap_perfo}
    \end{figure*}
\else
    \begin{figure*}[t]
    	\centering
    	\includegraphics[width=6.5in]{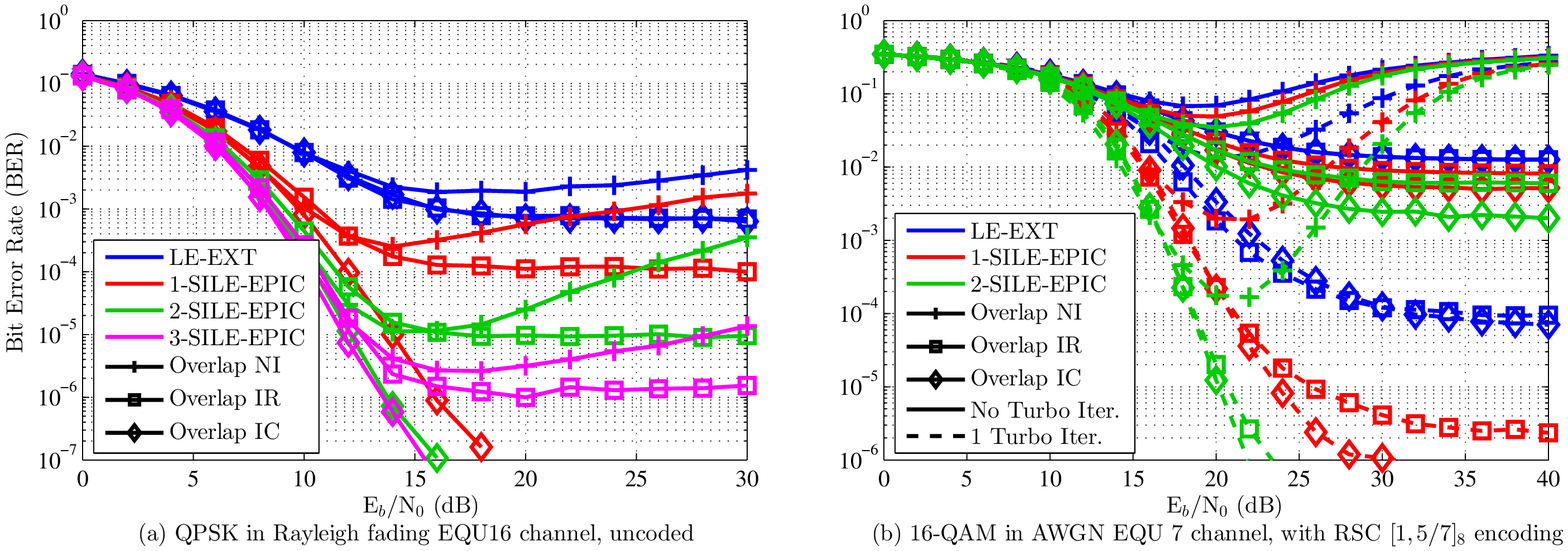}
    	\caption{IBI mitigation capabilities of overlap FDE schemes.}
    	\label{fig_overlap_perfo}
    \end{figure*}
\fi

\subsection{Overlap FDE with Interference Rejection}

Interference enhancement caused by overlap FDE NI causes prohibitive constraints for selecting $N$, $N_l$ and $N_r$, in order to avoid residual IBI. Moreover, for channel with severe spectral nulls, IBI spread can be as large as $N$, making overlap FDE unusable with any parameters. If the channel is also time-varying, coherent time constraints on $N$ are imposed, which may cause overlap FDE NI to yield no viable solution. 

In this paragraph, the interference rejection (IR) strategy, which mitigates interference enhancement, is exposed, by designing filters that account for the presence of IBI. 
The equivalent noise which also includes the IBI is
\begin{equation}
\tilde{\mathbf{w}}'_{k} \triangleq \mathbf{G}_k(\tilde{\mathbf{x}}_{k-N}-\tilde{\mathbf{x}}_{k}) + \tilde{\mathbf{w}}_{k}.
\end{equation}
Considering the noise model (see eq. (\ref{eq_circmodel})) used in the FDE design, in section \ref{sec:fdsile}, one can compute 
%$N$-symbol 
a SC-FDE equalizer 
(\ref{eq_filt1})-(\ref{eq_fde_outvar}) 
using the equivalent noise covariance
\begin{equation}
\mathbf{\underline{\Sigma}}_{\tilde{\mathbf{w}}_k'} = \mathbf{\underline{\Sigma}}_{\tilde{\mathbf{w}}_k} + 2\sigma_x^2\mathcal{F}_N\mathbf{G}_k\mathbf{G}_k^H\mathcal{F}_N^H, \label{eq_overlapIR_cov}
\end{equation}
assuming i.i.d. transmitted symbols.
The equalizer neglects noise correlations between different subcarriers, but accounts for the FD coloured noise with diagonals of matrix (\ref{eq_overlapIR_cov}). 
IR was applied using the whitened covariance of the IBI in \cite{tomeba2006overlap}, however using a coloured representation, as in this paper, was shown significantly improved performance \cite{obara_adachi_2001_mmseWeight}. This strategy does not suffer from error enhancement at high SNR, and produce steady error-floors. Nevertheless IR can perform slightly worse than NI at low SNR, due to pessimistic representation of IBI covariance.

\subsection{Overlap FDE with Interference Cancellation}

To completely remove residual IBI in overlap FDE with limited overlap interval, interference cancellation is needed, especially for highly selective channels, where equalization filter has time response of length comparable to FFT, and spreads IBI over all symbols. 

There are various approaches to IBI cancellation in overlap FDE, either with serial decision feedback for joint ISI/IBI cancellation \cite{tomasin_2005_overlapAndSaveFDDFE}, with hard decision feedback for successive IBI cancellation \cite{wangLiang_2008_FDEDownlinkSCDMA}, or with hybrid turbo and hard successive decision feedback \cite{ravaeiNejad_2017_LowComplexyfFDTEQHF}. 
Unlike these references, which uses decisions on previously processed sub-blocks, here we focus on parallel IBI cancellation, using solely a feedback generated from the previous SI/TI, for ensuring parallel processing of sub-blocks in practical implementations.
Moreover, EP-based feedback is used, as its overall superiority compared to EXT or APP feedback was shown in the previous section.

At $\tau=s=0$, IR is used via (\ref{eq_overlapIR_cov}), then IBI is removed before the $N$-point FFTs with
\begin{equation}
    \tilde{\mathbf{y}}'_{k} \triangleq \tilde{\mathbf{y}}_{k} - \mathbf{G}_k \left(\tilde{\mathbf{x}}^{d(\tau',s')}_{k-N}-\tilde{\mathbf{x}}^{d(\tau',s')}_{k} \right),
\end{equation}
where $\tau'$ and $s'$ denote the previous TI/SI index. 

Moreover, unlike prior work on overlap FDE-IC, we use adaptive IR, by accounting for the \emph{residual} IBI in filter weight computations with
\begin{equation}
\mathbf{\underline{\Sigma}}_{\tilde{\mathbf{w}}_k'} = \mathbf{\underline{\Sigma}}_{\tilde{\mathbf{w}}_k} + 2{v}^{d(\tau',s')}\mathcal{F}_N\mathbf{G}_k\mathbf{G}_k^H\mathcal{F}_N^H.
\end{equation}
As in overlap FDE NI/IR strategies above, $N_b$ parallel equalizers are operated concurrently for detecting all sub-blocks. Finally, it is possible, depending on the channel coherence time, to set $N_l=N_r=0$, for $\tau>0$, to reduce $N_b$, as in \cite{ravaeiNejad_2017_LowComplexyfFDTEQHF}, to reduce the receiver complexity.

\subsection{Inter-block interference mitigation performance}

In this section, $K$-block quasi-static channels are considered, to focus on the EP-based overlap FDEs' IBI mitigation capabilities.
The benefits of SI are compared to the conventional FD LE-EXTIC (i.e. $S=0$), for overlap FDE, possibly equipped with IR and/or IC. The IC strategy of setting $N_l=N_r=0$ for $\tau>0$ for overlap FDE IC is used for these simulations.

First we consider an uncoded scenario, similar to the benchmark \cite{obara_adachi_2001_mmseWeight}, with QPSK constellation in a quasi-static Rayleigh fading frequency-selective channel with symbol spaced 16-path uniform power delay profile (EQU16). Transmission parameters are $K=2048$, $N=256$ and $N_l=N_r=16$, and 80000 block transmissions per SNR are used to numerically approximate the BER for $\mathcal{S}=0\dots3$ in Fig. \ref{fig_overlap_perfo}-(a). 
The conventional scheme (NI) is unusable, as the overlap interval is insufficient to contain all IBI, and SI ($\beta=0.25\times 0.5^{s+\tau}$) do not resist to IBI amplifications. But IR significantly benefits from SI, as it further reduces the error floor.
Finally, overlap IC with SIs removes most of the interference, even with a single SI.

A more extreme case, with strong IBI, is considered in Fig. \ref{fig_overlap_perfo}-(b) (16-QAM, RSC $[1,5/7]_8$), within a 7-path static AWGN channel, with uniform power delay profile.
We consider 50000 block transmissions with $K=1024$, $N=128$, and $N_l=N_r=7$, to evaluate the BER. In this case, SI ($\beta=0.75\times 0.9^{s+\tau}$) alone cannot remove error floors even with IC and channel coding, but with the help of a single TI, even E IR's error floor, with EP-based SI, becomes at least two order of magnitudes smaller than traditional FD LE-EXTIC.

\subsection{Performance in a doubly-selective channel}

\ifdouble
    \begin{figure*}[t]
    	\centering
    	\includegraphics[width=7in]{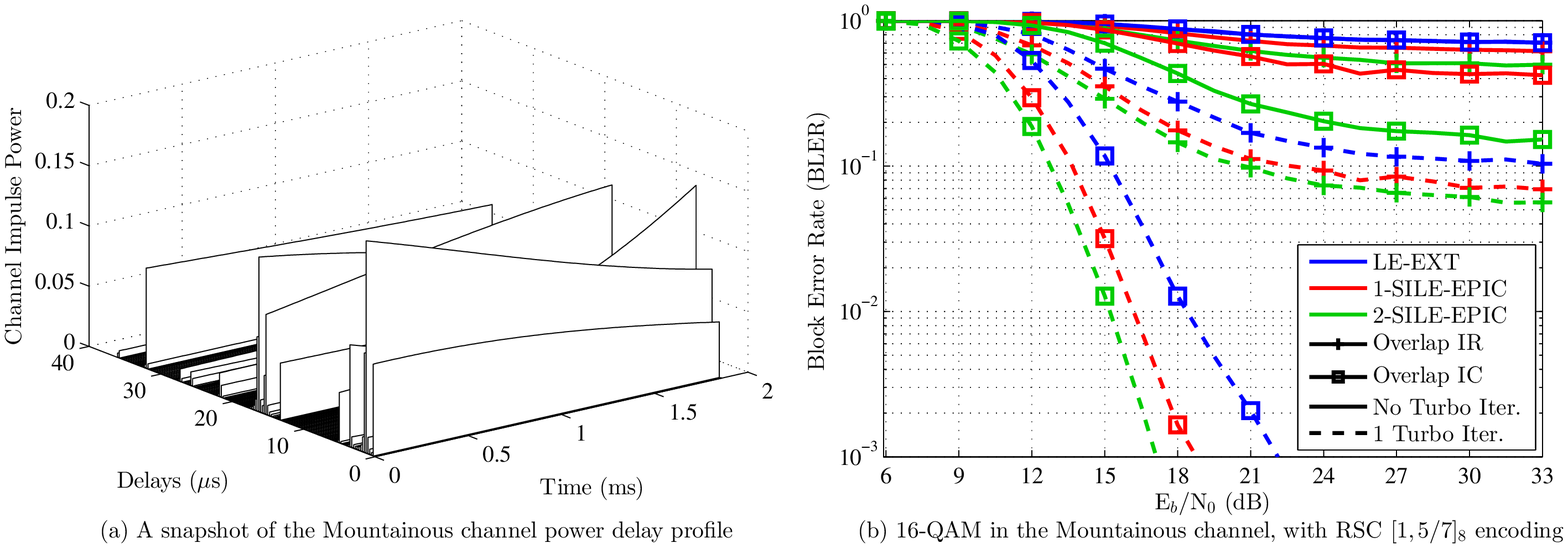}
    	\caption{EP-based overlap FDE performance in the doubly-selective mountainous channel.}
    	\label{fig_overlap_perfo_tv}
    \end{figure*}
\else
    \begin{figure*}[t]
    	\centering
    	\includegraphics[width=6.5in]{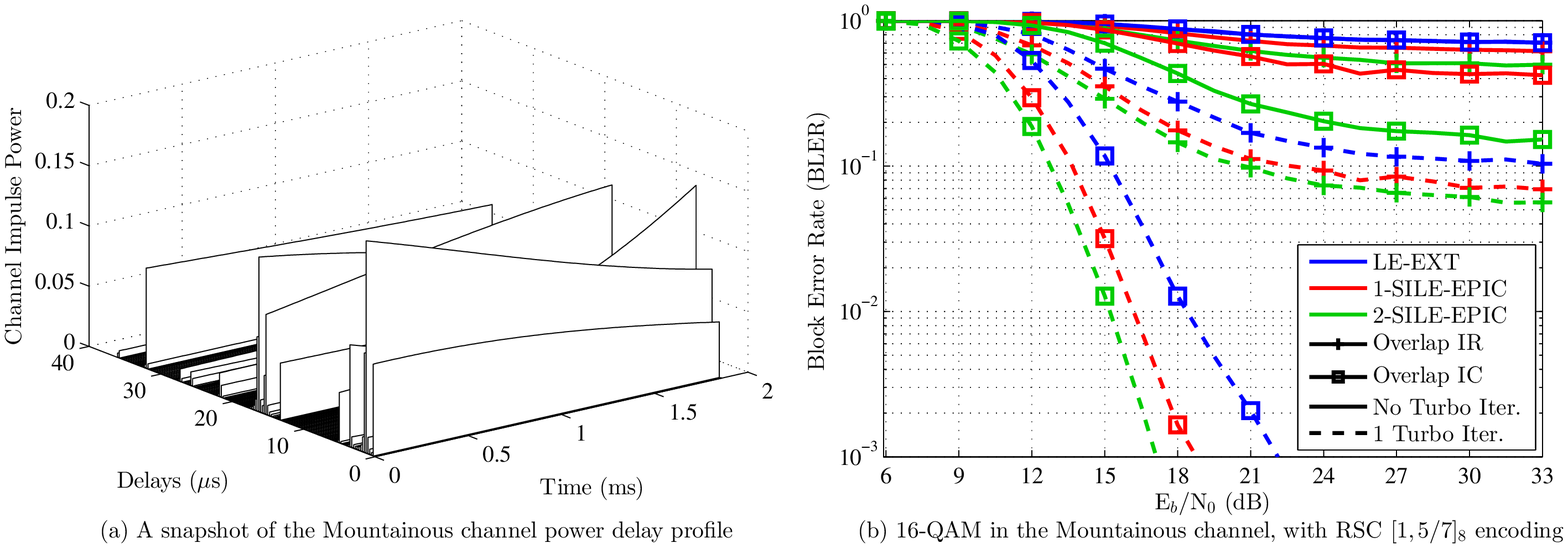}
    	\caption{EP-based overlap FDE performance in the doubly-selective mountainous channel.}
    	\label{fig_overlap_perfo_tv}
    \end{figure*}
\fi

The behaviour of the overlap FDE with the proposed EP-based self-iterations is evaluated within a mobile ad-hoc network (MANET) scenario where mobile-to-mobile communications between two high-speed vehicles is considered in a harsh environment. The mountainous channel model from \cite[Tab. 5.10]{fischer_2017_PHYLinkModelMANET} is used. Vehicles are assumed to move at 130~km/h each, in opposing directions, hence generating a maximum Doppler shift of 96~Hz, assuming the use of a carrier frequency at 400~MHz. A snapshot of a random channel realization is plotted in Fig. \ref{fig_overlap_perfo_tv}-(a).

SC transmissions with 1/2-rate-coded 16-QAM constellation is considered, with a baud-rate of 1~Mbauds/s, and a root raised-cosine pulse-shaping with a roll-off factor of 0.35. In this case, the base-band channel spread is $L=45$ symbols. $N=256$ symbol is chosen to ensure that the channel remains static on each sub-block. We consider $K=1536$ and $N_l=N_r=18$, an ovelap length of 18 symbols is chosen as most significant paths of the mountainous channel, (and other urban, hilly or rural channels in \cite{fischer_2017_PHYLinkModelMANET})  are contained within 18~$\mu s$.

In Fig. \ref{fig_overlap_perfo_tv}-(b), the block error rate (BLER) of overlap FDE IR/IC are plotted. It can be seen that IR cannot get rid of the error floor but using overlap IC and one TI, robust transmissions are possible. In this case, one and two SIs ($\beta=\min(0.5,0.7^{1+s+\tau})$) respectively bring 2.7~dB and 3.9~dB improvements, at BLER = $3.10^{-3}$.
The use of SC-FDE with six block transmissions of $K=256$, with cyclic prefix and guard intervals to avoid IBI, instead of using the considered overlap FDE, would have required 90 additional symbol slots per block, and would have caused a loss of throughput and energy-efficiency of respectively 12~\% and 0.6~dB.

\section{Application to Multi-Antenna Spatial Multiplexing: FD MIMO Receiver}\label{sec:fdMIMO}
\subsection{System Model and Overview of Resolution}

Here, the extension of the SC-FDE model in section \ref{sec:modelling}, to incorporate multiple antennas is considered. The transmitter and the receiver have respectively $T$ and $R$ antennas, and space-time bit-interleaved coded modulation (STBICM) is used \cite{tonello2000space}. This ensures the transmitted symbols blocks $\mathbf{x}_t$ on each transmit antenna $t$ to be independent of each other, and coded bits $\mathbf{d}_{k,t}\in \mathbb{F}_2^q$ associated to each symbol to be bit-wise independent as a generalization of the BICM.

Assuming the use of a CP on each antenna, and using ideal synchronization and ideal channel state knowledge hypotheses at the receiver, received samples on the $r^\text{th}$ antenna are
\begin{equation}
	\mathbf{y}_r = \textstyle \sum_{t=1}^{T}\mathbf{H}_{r,t}\mathbf{x}_{t} + \mathbf{w}_{r},
\end{equation}
where the $K \times K$ matrix $\mathbf{H}_{r,t}$ is the circulant channel matrix associated to the $L$-tap impulse response $[h_{1,r,t},\dots,h_{L,r,t}]$ of the channel between $t^\text{th}$ TX, and the $r^\text{th}$ RX antennas, and where the noise $\mathbf{w}_r ~\sim~ \mathcal{CN}(0, \sigma_w^2 \mathbf{I}_N)$. 
The FD channel is $\mathbf{\underline{H}}_{r,t}=\mathcal{F}_K\mathbf{H}_{r,t,u}\mathcal{F}_K^H$, as in section \ref{sec:modelling}.
Stacking receiver antennas to form $\mathbf{\underline{y}}=[\mathbf{\underline{y}}_1;\dots;\mathbf{\underline{y}}_{R}]$,
transmit antennas for $\mathbf{{x}}=[\mathbf{{x}}_{1};\dots;\mathbf{{x}}_{T}]$, we have
\begin{equation}
	\mathbf{\underline{y}} = \mathbf{\underline{H}}\mathcal{F}_{K,T}\mathbf{{x}} + \mathbf{\underline{w}},
\end{equation}
where $\mathbf{\underline{H}}$ is a $RK\times TK$ $K$-partitioned-diagonal matrix, 
$\mathcal{F}_{K,T}=\mathbf{I}_T\otimes \mathcal{F}_{K}$ is the $T$-block DFT matrix, and $\mathbf{\underline{w}}=[\mathbf{\underline{w}}_1;\dots;\mathbf{\underline{w}}_{R}]$ with $\mathbf{\underline{w}} \sim \mathcal{CN}(\mathbf{0}_{RK}, \sigma_w^2 \mathbf{I}_{RK})$.

\ifdouble
\begin{figure*}[t]
	\centering
	\includegraphics[width=7in]{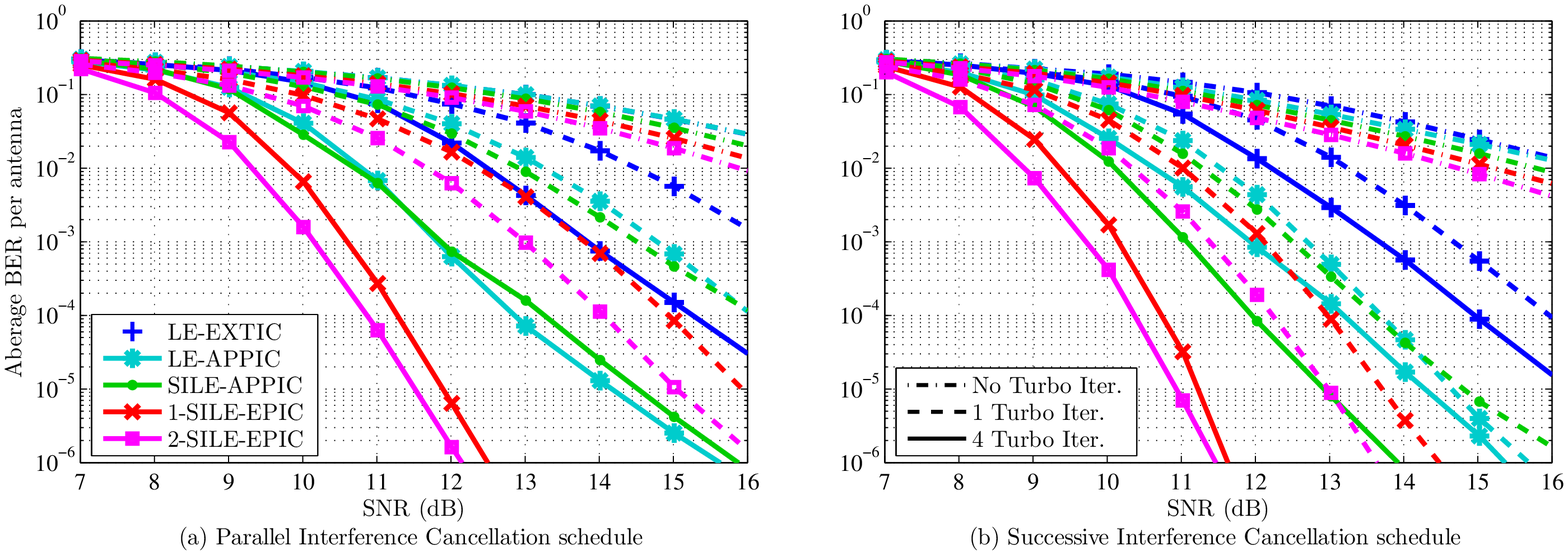}
	\caption{BER in $2\times 2$ MIMO spatial multiplexing in the Proakis B with 16-QAM, using rate-1/2 $[17, 13]_8$ convolutional code.}
	\label{fig_ber_mud_proakisb}
\end{figure*}
\else
\begin{figure*}[t]
	\centering
	\includegraphics[width=6.5in]{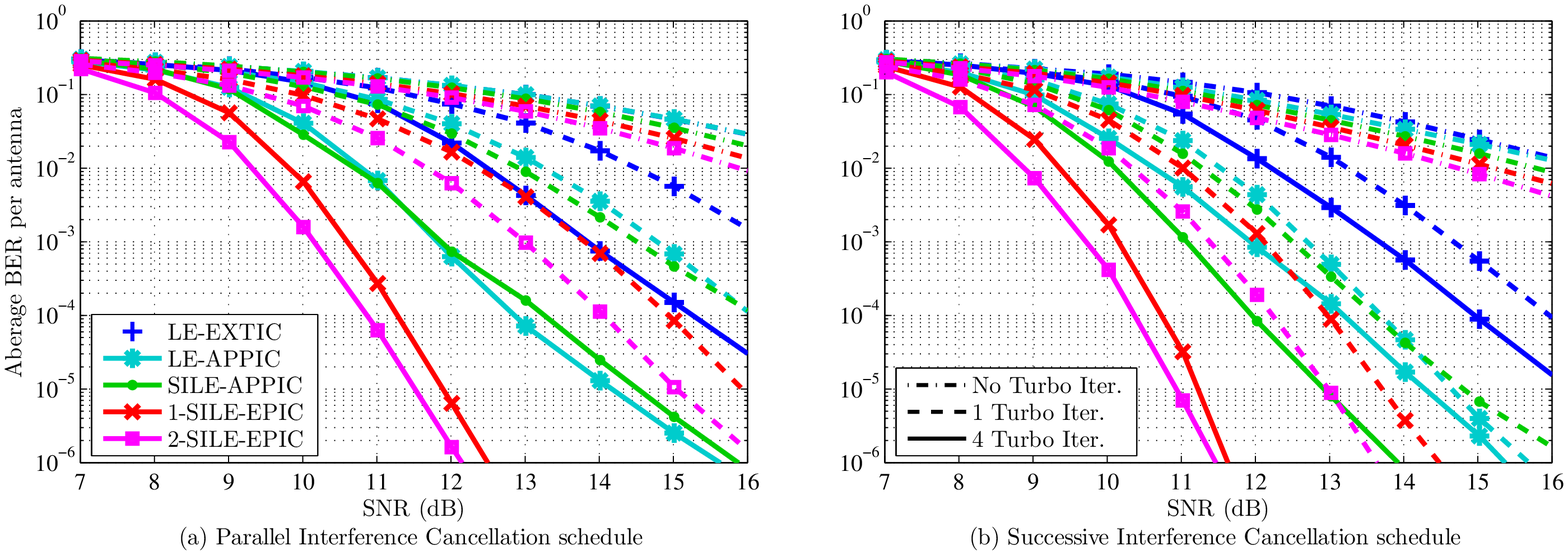}
	\caption{BER in $2\times 2$ MIMO spatial multiplexing in the Proakis B with 16-QAM, using rate-1/2 $[17, 13]_8$ convolutional code.}
	\label{fig_ber_mud_proakisb}
\end{figure*}
\fi

A MIMO detector can be designed by applying the proposed framework on the joint PDF $p(\mathbf{d}, \mathbf{x}  \vert \mathbf{\underline{y}})$ of this STBICM system, factorized as $p(\mathbf{\underline{y}} \vert \mathbf{x})\prod_{t=1}^{T}\prod_{k=0}^{K-1}p(x_{k,t}\vert \mathbf{{d}}_{k,t})\prod_{j=0}^{q-1} p(d_{k,j,t})$.
Detailed derivation is not given here due to lack of space, however  a multi-user MIMO system for non-orthogonal multiple-access with SC-FDMA waveform, which generalizes this system, is derived in \cite{sahin_2018_SpectrallyMUDEP}. 
The resulting MIMO detector's outputs are given by
\begin{align}
	\underline{x}^{e}_{k,t} &= \underline{x}^{d}_{k,t} +
	\textstyle\sum_{r=1}^R \underline{f}_{k,r,t}^{*} \, (\underline{y}_{k,r} - \textstyle\sum_{t=1}^{T} {\underline{h}}_{k,r,t}\underline{x}^{d}_{k,t}), \label{eq_EQU_extsym}\\
    {v}^{e}_{t} &=   1/\bar{\xi}_{t} - {v}^{d}_{t}, \label{eq_EQU_extvar}
\end{align}
where $\underline{f}_{k,r,t} = \bar{\xi}_{t}^{-1} \textstyle\sum_{r'=1}^R \lambda^{d}_{k,r,r'} {\underline{h}}_{k,r',t}$, with $\lambda^{d}_{k,r,r'}$ being the $k^\text{th}$ diagonal of $\mathbf{\Sigma}^{\mathbf{d}}{}^{-1}$'s  $(r,r')^\text{th}$ partition,
and $\bar{\xi}_{t} = K^{-1} \textstyle\sum_{r=1}^R {\underline{h}}_{k,r,t}^{*} \sum_{r'=1}^R \lambda^{d}_{k,r,r'} {\underline{h}}_{k,r',t}$. The covariance matrix $\mathbf{\Sigma}^{\mathbf{d}}$ is given by
\begin{equation}
	\mathbf{\Sigma}^{\mathbf{d}} = \textstyle \sigma_w^2\mathbf{I}_{RK} + \sum_{t=1}^{T} {v}^{d}_{t}\mathbf{\underline{H}}_{t}\mathbf{\underline{H}}_{t}^H, \label{eq_MUDCovMtxRecursive}
\end{equation}
where $\mathbf{\underline{H}}_{t}\in \mathbb{C}^{RK\times K}$ is given by the partitioning  $\mathbf{\underline{H}} =$ $\left[\mathbf{\underline{H}}_{1,1},\dots,\mathbf{\underline{H}}_{T_1,1},\dots,\mathbf{\underline{H}}_{T_U,U}\right]$. 
This covariance matrix and its inverse have a partitioned-diagonal structure, which allows using $\lambda^{d}_{k,r,r'}$ for computationally-efficient detection. For each antenna $t$, a separate EP-based demapper is used, with their specific input and output variances, i.e. respectively $v^e_t$ and $v^d_t$, for characterizing temporally white soft estimates of transmitted symbols.

\subsection{Performance Comparison}

The proposed FD MIMO detector is evaluated in the spatial-multiplexing scenario of \cite{tao_2015_singlecarrierfreq}; over the generalized AWGN Proakis B channel with $T=R=2$, $K=128$. Up to 2~SIs are considered with $\beta=\max(0.3,0.5\times(0.8)^{s+\tau})$, and average BER per antenna is evaluated.
In Fig. \ref{fig_ber_mud_proakisb}-(a), the multi-antenna interference (MAI) is mitigated with a parallel IC (PIC) schedule, as in \cite{tao_2015_singlecarrierfreq}, i.e. with simultaneous detection over antennas in each SI, and simultaneous decoding of all antennas in each TI. Our proposal displays remarkable gains over APP-based prior work, with over 2~dB and 2.5~dB gains at 4~TIs, at $\text{BER}=10^{-5}$, with respectively 1 and 2~SIs.

In Fig. \ref{fig_ber_mud_proakisb}-(b), the MAI is mitigated with a successive IC (SIC) schedule, as in \cite{sahin_2018_SpectrallyMUDEP}, i.e. with simultaneous detection over all antennas in each SI, but with successive decoding of antennas in each TI. This approach is known to converge faster. 
Our proposal outperforms concurrent structures for all TI, with over 1.5~dB margin for $\text{BER}=10^{-5}$. Moreover, \emph{SILE-APPIC} with either SIC or PIC, at 4~TIs, is outperformed by either \emph{1-SILE-EPIC} with $\mathcal{T}=1$ with PIC or SIC. Asymptotically ($\mathcal{T}=4$), SIC improves our proposal's BER around 0.5~dB over PIC, but SIC with 1 TIs is shown to significantly outperform alternatives, which provides an attractive compromise of fewer decoder iterations, but increased detector iterations, to provide an attractive complexity-performance options, especially when using powerful decoders.

\section{Conclusion}\label{sec:concl}

This paper considers a new framework based on expectation propagation message passing for the design of low-complexity digital receivers. This approach's particularity lies in constraining transmitted symbol variable nodes to lie in multivariate temporally-white Gaussian distributions. This allows deriving low-complexity EP-based demappers which can provide extrinsic symbol-wise feedback, whose reliability measure is characterized by a scalar variance.

The proposed methodology is exposed through the design and analysis of an elementary SC-FDE receiver, which is shown to be either more energy-efficient or less complex than alternatives of the state of the art.
Resulting receiver can be seen as a double-loop, low-complexity single-tap FDE which can achieve remarkable energy savings with conventional forward error correction techniques.
In particular, it is shown through asymptotic analysis, that a considerable portion of the channel symmetric information rate region becomes achievable.

Furthermore, the flexibility of the proposed approach for receiver design is shown by applying it to the three categories of overlap FDE receivers, and a MIMO detector for spatial multiplexing. In all cases, significant improvements were observed in terms of performance-complexity trade-off. 
Other practical applications of this framework will be exposed in future works.

%\appendix

%\subsection{Derivation of ...}

\ifCLASSOPTIONcaptionsoff
  \newpage
\fi



%\bibliographystyle{IEEEtran}

\bibliographystyle{IEEEtran}
\balance
\bibliography{IEEEabrv,bibDFE}

\end{document}